Exploratory studies of human gait changes using depth cameras
and sample entropy

By

Behnam Malmir

A THESIS

Submitted in partial fulfillment of the requirements for the degree

MASTER OF SCIENCE

Department of Industrial and Manufacturing Systems Engineering
College of Engineering

KANSAS STATE UNIVERSITY
Manhattan, Kansas

2018


Approved by:

Major Professor
Shing I Chang


# Abstract

This research aims to quantify human walking patterns through depth cameras to (1) detect walking pattern changes of a person with and without a motion-restricting device or a walking aid, and to (2) identify distinct walking patterns from different persons of similar physical attributes. Microsoft Kinect™ devices, often used for video games, were used to provide and track coordinates of 25 different joints of people over time to form a human skeleton.

Two main studies were conducted. The first study aims at deciding whether motion-restricted devices such as a knee brace, an ankle brace, or walking aids – walkers or canes affect a person's walking pattern or not. This study collects gait data from ten healthy subjects consisting of five females and five males walking a 10-foot path multiple times with and without motion-restricting devices. Their walking patterns were recorded in a form of time series via two Microsoft Kinect™ devices through frontal and sagittal planes. Two types of statistics were generated for analytic purposes. The first type is gait parameters converted from Microsoft Kinect™ coordinates of six selected joints. Then Sample Entropy (SE) measures were computed from the gait parameter values over time. The second method, on the other hand, applies the SE computations directly on the raw data derived from Microsoft Kinect™ devices in terms of (X, Y, Z) coordinates of 15 selected joints over time. The SE values were then used to compare the changes in each joint with and without motion-restricting devices. The experimental results show that both types of statistics are capable of detecting differences in walking patterns with and without motion-restricting devices for all ten subjects.

The second study focuses on distinguishing two healthy persons with similar physical conditions. SE values from three gait parameters were used to distinguish one person from another via their walking patterns. The experimental results show that the proposed method using a star glyph summarizing the shape produced by the gait parameters is capable of distinguishing these two persons.

Then multiple machine learning (ML) models were applied to the SE datasets from ten college age subjects - five males and five females. In particular, ML models were applied to classify subjects into two categories: normal walking and abnormal walking (i.e. with motion-restricting devices). The best ML model (K-nearest neighborhood) was able to predict 97.3% accuracy using 10-fold cross-validation. Finally, ML models were applied to classify five gait conditions: walking normally, walking while wearing the ankle brace, walking while wearing the ACL brace, walking while using a cane, and walking while using a walker. The best ML model was again the K-nearest neighborhood performing at 98.7% accuracy rate.

# Table of Contents







# List of Figures









# List of Tables





# List of Equations





# Acknowledgments

I would like to express my deepest gratitude to my advisor, Dr. Shing Chang for guiding and supporting me over past three years. You have set an example of excellence as a researcher, mentor, teacher, and a friend. I am thankful to my co-supervisor, Dr. Malgorzata Rys for her valuable guidance, supports, and feedbacks. I would like to thank my thesis committee members for all their guidance through this process; your discussion, ideas, and feedback have been undeniably invaluable. I would like to thank my fellow undergraduates who contributed to this research.

I am very grateful every day for my amazing family and their eternal love, support, and constant enthusiasm and encouragement which have always kept me pleased and inspired.

The thesis is dedicated to the most remarkable person I know, my extraordinary mother who played an incredible role in my life. I am grateful for your love, kindness, wisdom, and sacrifices. I am very blessed to have you and I owe you every single success I hope to obtain.

"You can never be overdressed or overeducated."
— **<u>Oscar Wilde</u>**



# Chapter 1.  Introduction

## 1.1. Background

Human gaits can be used as an important indicator of health in a wide range of diseases, such as

diabetes (Hodgins, 2008), neurological diseases (Keijsers et al., 2006), (Hausdorff et al., 2000),

and fall detection and prediction (Hausdorff et al., 2001). Many researchers have used these

indicators in their studies. Welsh et al. (2018) used some gait parameters such as step width,

length, and other gait measurements as gait indicators to analyze changes in stability. Their study

relates the changes observed in the stability to overall views of health. MacDonald et al. (2017)

used timed walks and an analysis of gait parameters to predict changes in a patient's cognitive

function. Their study showed that a timed walk is an effective tool for predicting mental changes.

Terrier and Reynard (2015) provided an in-depth look at how age affects the gait. The information

can be used to better understand how health and gait are connected. Another study conducted

by Verlinden et al. (2015) proposed that gait becomes different based on gender when a person

grows old. This information can also be used to help understand how gait changes with age. Smith

et al. (2017) found that speed changes in gaits often occurred when the mind was occupied. Their

study could help prove that gait patterns may serve as a good indicator of mental health and

wellness. In another study, Eggleston et al. (2017) also connected gait to mental wellness,

specifically aimed at the mentally ill. Raknim and Lan (2016) used smartphone sensors to analyze

a person's gait and examined a test to see if smartphones might be connected to neurological

disorders. This study further proves the connection between gait and overall health. In addition,

Esser et al.'s (2018) article related the decreased freedom that some diabetes patients have due



to a neurological issue connected to diabetes. They proposed their own method and technology for monitoring human gait.

Ayoubi et al. (2015) analyzed how the fear of falling contributes to gait variability in the elderly population. The findings of this study showed that a person's gait can change solely based on their fear of falling. Having this knowledge is valuable in understanding the neurological effects on gait. Lye et al.'s (2015) research looked at gait differences in young children, specifically how the subjects propel themselves during their gait cycle. The results showed how underutilization of ankle power suggest physical immaturity and may be an indicator of other health issues. Bahl et al. (2018) focused mainly on rehabilitation type situations after a hip replacement. This study showed how gait tracking can be used to track progress in rehabilitation. Moreover, Christensen et al.'s (2018) looked at gait changes after a knee replacement and analyze the differences between the test and healthy control groups. In addition, Zou et al. (2017) surveyed the effects of a healthy lifestyle on an elderly person's gait. It showed that keeping elderlies moving is a key to prevent falls.

Muro-De-La-Herran et al. (2014) published an overview of some gait analysis methods to better understand the effectiveness and accuracy of each method. The results provided several more gait parameters which can be used for gait analysis. Finally, Khandelwal (2018) studied several important real-world applications of gait monitoring and their usefulness in predicting overall health.

The main motivation for the current research is to see whether motion-restricting devices alter the walking patterns of healthy subjects, and then try to detect and quantify the possible changes



in their walking patterns using some statistical and machine learning (ML) techniques. In addition, if gaits can be quantified by (X, Y, Z) coordinates from multiple joints using a depth camera, can gait from different people be distinguished through statistics generated from these joints?

## 1.2. Problem Statements

This study aims to quantify and detect human gait changes. The proposed method summarizes walking patterns using human skeleton coordinates recorded and derived from Microsoft Kinect™ devices over time. The proposed procedure can be applied to fall prediction of fall of elderly people, physical therapy, and sports science.

In this study, we also study an approach for gait recognition based on human skeleton coordinates. Microsoft Kinect™ with an integrated depth sensor, enabling skeleton detection and tracking in real-time, has been used in this approach. The proposed approach is expected to distinguish different people through their normal gait patterns.

Similar studies have already shown the potential of Microsoft Kinect™ for fall risk assessment (Rantz et al., 2015), as well as a clinical and field-based assessment of gait (Mentiplay et al., 2015; Procházka et al., 2015), such as Timed Up and Go (TUG). Mousavi Hondori and Khademi (2014) studied the technical and clinical impacts of Kinect™ on physical therapy and rehabilitation. The subjects considered in this research include elderly patients with neurological disorders, strokes, Parkinson, cerebral palsy, and MS. Other issues, such as sleep disorders (Centonze et al., 2015) or recognition of breathing (Schatz et al., 2015), have been also analyzed using Kinect™. However, the current research tries to evaluate its abilities in detecting different walking patterns and discriminating an abnormal walking pattern from an abnormal one.



## 1.3. Proposed Procedure

Below are the main steps of the proposed method and data analysis:

1. Set up an experimental protocol containing Kinect™ set-up information such as camera angles, height and location, measured path and experimental layout for the test, and a checklist to make sure that all required conditions such as particular clothes, shoes, their colors, and light of the location are met prior to starting a test.

2. Recruit subjects and set up the test equipment and the environment using the protocol. Two operators were recruited to monitor and record people's walks using two computers connected to depth cameras and a group of subjects voluntarily participated in this study. A questionnaire was provided to collect information related to age, height, and shoe size.

3. Collect subjects' gait data by recording their selected joints' coordinates, as the indicators of their physical health. The test subjects were instructed to walk with a certain speed, lead with the same foot for each walk, and constantly wear similar fitting clothing and athletic shoes to avoid interfering with the results of the experiment.

4. Extract each test result from the computers and save them to a text file. Each file contains a time stamp, (X, Y, Z) coordinates and joint types. These coordinates were converted into gait parameters such as spine tilt, hip tilt, and shoulder tilt when data is collected from the front camera. Each gait parameter or joint along with their coordinates recorded over time makes a time series.

5. Use Sample Entropy (Richman and Moorman, 2000) to convert each time series to a numerical value. SE values serve as indicators for gait variabilities and changes over time.



6. Analyze gaits from the SE measures of either gait parameter or raw data. Everyone has their own walking patterns (Kuchler, 2017; Palastanga et al., 2012). Thus, we tag subjects' test data derived from Microsoft Kinect™ cameras as the training data. Machine learning algorithms were used to compare and classify these SE indicators in detecting gait changes or distinguishing different persons.

## 1.4. Thesis Outline

This thesis contains the following chapters. Chapter 1 defines the research problems and outlines the proposed method along with a brief literature review of all methods and techniques used to analyze human gait to date. Chapter 2 outlines the proposed integrated system, its operation, and its accuracy for analyzing human gait. Chapter 3 describes the experimental plan of the study containing participants, equipment, and set up processing. It also explains the raw data type, gait parameters and visualizes them. Chapter 4 contains all experiments conducted for two main studies – a detecting gait changes and identifying different persons. It provides examples of statistical analyses, machine learning predictions, and visualization tools. Chapter 5 includes a summary of the research, conclusions and future studies.



# Chapter 2.  Setup of Depth Cameras and Accuracy Studies

This chapter is divided into three subsections. The first subsection introduces the best device among 3D motion capture systems in terms of accuracy, reliability, and cost for gait analysis. The second subsection simply explains how this system operates, and the third subsection examines the accuracy of the proposed system for analyzing human gaits.

## 2.1. Why Microsoft Kinect™?

As discussed in chapter one, three-dimensional (3D) motion capture systems are mostly used (Pfister et al., 2014) in kinematic analysis and studies on human gait. However, most of them are costly and not available in clinical settings. The Microsoft Kinect™ has the potential to record joint coordinates of the human skeleton in three-dimensional (3D) space. It is an accurate, non-intrusive, low-cost clinical gait analysis system that has many applications in diagnosis, monitoring, treatment and rehabilitation (Wren et al., 2011; Hodgins, 2008). Such applications include early diagnosis and assessment (Hoff et al., 2001), measuring medication effectiveness at home (Keijsers et al., 2003), and even direct treatment optimization (Legros et al., 2004; Holt et al., 2011). Researchers have widely used these 3D kinematic measures in clinical gait analysis studies since 2011, although there are some other related articles published even before 2011. Figure 2.1 demonstrates how fast this area of research has expanded.

For instance, Tupa et al. (2015) used Kinect to estimate leg length, normalized average stride length, and gait velocity of an individual. These gait features were then compared in three sets of individuals to recognize Parkinson's disease.



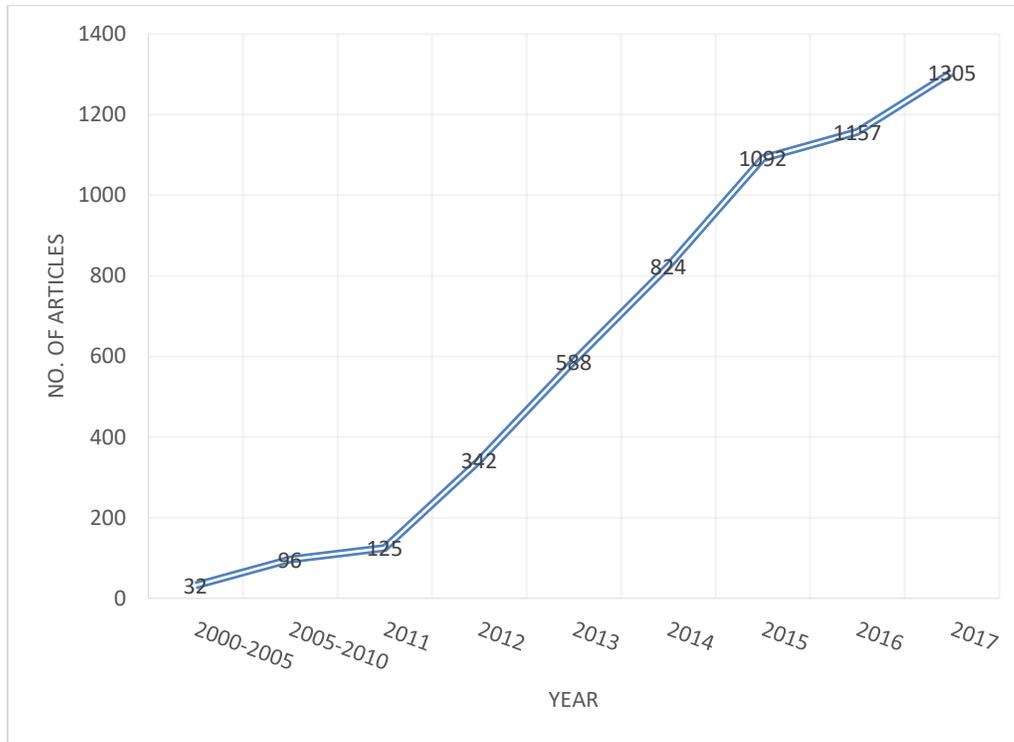

**Figure 2.1 Diagram of using Microsoft Kinect™ in gait studies**

Vernon et al. (2015) also examined the test-retest reliability measures of some other kinematic measures, such as step length and stride length, to determine whether they can improve prediction performance in common clinical tests. Eltoukhy et al. (2017) used Kinect to identify different patients. Their study showed gait patterns differed between healthy people and those with Parkinson's disease. Xu et al. (2017) tracked shoulder movements while a person used a computer in an effort to reduce injuries. Their study found placing Kinect to the front of participants yielded more accurate shoulder measurements than placing the camera 15 to 30 degrees to the side. Auvinet et al. (2015) used Kinect to detect gait cycles through detecting the actual heel-strike event. They demonstrated that there is a relationship between maximum DK (Distance between knees along the longitudinal walking axis) values and heel-strike events in the gaits of healthy subjects.



## 2.2. Kinect™ Operation

Microsoft Kinect$^{TM}$ was initially designed to enhance the video gaming experience by capturing a gamer's joint position. Consequently, Kinect presents a simple, inexpensive, and portable method of examining motion in a human subject test, such as Timed Up and Go (TUG), without intrusion on human subjects (Vernon et al., 2015).

Kinect is unique and useful for gait analysis because it contains an RGB camera, a depth sensor, and a multi-array microphone. Kinect's depth sensor can capture 3D data and does not require any particular lighting for the system, allowing it to capture data indoors or outdoors. In this study, however, we used the same room with the same lighting throughout the experimental period to eliminate the possibilities of bias data collections. In the current study, Kinect was set to record 30 fps of all joints in two directions, as shown in Figure 2.1. However, Kinect is capable of a frame rate of 9–30 fps and a resolution of 640 x 480 that can be increased to 1280 x 1024 using a lower frame rate (Prochazka et al., 2015). A customized software based on the software development kit (SDK) was developed in a C# language to gather data under a Windows operation system. A dynamic link library (DLL) was used to obtain the coordinates of selected skeleton joints in three different axes (X, Y, Z) = (anteroposterior, vertical, and mediolateral). Figure 2.1 illustrates an example of 3D capture for the neck, drawn in Matlab R2014a.



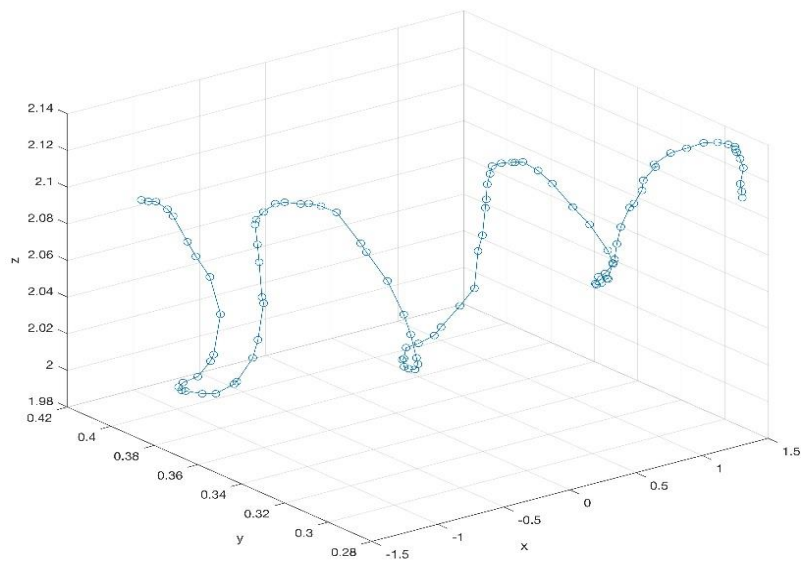

**Figure 2.2 3D coordinates of the neck acquired from the Microsoft Kinect™ over time**

## 2.3. The accuracy of Microsoft Kinect™ for Analyzing Gait

As discussed in section 2.2, Kinect helps examine motion in a human subject test without intruding on the subject through a simple, inexpensive, and portable method (Vernon et al., 2015). Many studies have found it unique and useful for gait analysis. The effectiveness of Kinect for gait analysis was tested in the study by Tupa et al. (2015), achieving an accuracy of 97.2% and suggesting the potential use of Kinect image and depth sensors for many applications. However, the accuracy level using the Kinect sensors differs across various joints and gait parameters (Kharazi et al., 2015). Xu and McGorry (2015) examined the accuracy of the Kinect sensor-identified coordinates of joint locations during 8 standing and 8 sitting postures of daily activities. The results of their study indicated their proposed alignment method can successfully align the Kinect sensor with respect to the motion tracking system. Their study further examined the



accuracy level of the Kinect sensor for the assessment of various gait parameters during treadmill walking under different walking speeds (Xu et al., 2015).

Some other studies compared Microsoft Kinect[TM] to other motion capture systems. For instance, Clark et al. (2012) assessed the concurrent validity of the Microsoft Kinect[TM] against a multiple-camera 3D motion analysis system during three postural control tests: forward reach, lateral reach, and single-leg eyes-closed standing balance. The Microsoft Kinect and 3D motion analysis systems had comparable inter-trial reliability and excellent concurrent validity, with Pearson's r-values > 0.90 for the majority of measurements. Clark et al. (2013) further assessed the concurrent validity of overground walking spatiotemporal data recorded using a marker-based three-dimensional motion analysis (3DMA) system and the Microsoft Kinect™. The outcome measures of gait speed, step length and time, stride length and time and peak foot swing velocity were derived using supervised automated analysis on twenty-one healthy adults who performed normal walking trials while being monitored by both systems. Gait speed, step length and stride length from the two devices possessed excellent agreement.

Clark et al.'s (2015) study also assessed the concurrent validity and reliability of kinematic data recorded using a marker-based 3DMA system and the Kinect V2 during a variety of static and dynamic balance assessments. Task-specific outcome measures from each system on Day 1 and 2 were compared. Concurrent validity of trunk angle data during the dynamic tasks and anterior-posterior range and path length in the static balance tasks was good (Pearson's r > 0.75). So, they claimed that Kinect V2 would have the potential to be used as a reliable and valid tool for the assessment of some aspects of balance performance. Thus, these findings suggest that the Microsoft Kinect can realistically assess kinematic strategies of postural control.



The accuracy of the Microsoft Kinect sensor for measuring clinically relevant movements is also established in people with Parkinson's disease (PD) (Galna et al., 2014). They used a Vicon three-dimensional motion analysis system (gold-standard) and the Microsoft Kinect concurrently to measure a series of movements performed by nine people with PD and ten controls. Their results from the Kinect related strongly to those obtained with the Vicon system (Pearson's r > 0.8) for most movements. Furthermore, the performance of the Microsoft Kinect sensors was compared with a force plate in Yeung et al.'s (2014) study. They concluded that the Kinect device was more accurate in the medial-lateral direction than in the anterior-posterior direction. The Kinect device also performed better than the force plate in more challenging balance tasks.

Some potential drawbacks are also declared in some studies. For example, concurrent validity for medial-lateral range and path length in Clark et al.'s (2015) study was poor to modest for all trials except one type of challenging balances. Moreover, test-retest reliability of a device was not consistent. However, the results were generally comparable between devices. Clark et al. (2012) concluded that there are proportional biases for some outcome measures, meaning that an increasing magnitude of the measure tends to broaden the difference between Microsoft Kinect measurements and those captured by the 3D camera system. Another study by Clark et al. (2013) also warned researchers about choosing appropriate and measurable outcome variables as some commonly reported variables in previous studies cannot be accurately measured by Kinect.

The Analysis of Variance (ANOVA) method (Houf, 1988) is applied in the following section to assure the reliability of the depth cameras used in the proposed studies reported in this thesis.



### 2.3.1. Gage Capability Study on Microsoft Kinect™ for Gait Analysis

In this section, the gauge R&R study is designed to assess the validity of the Kinect for analyzing gait of a 10-foot walk. The gauge R&R aims to estimate various components of measurement system variability. There are three main factors causing the variations: different days, different trials, and different subjects. Table 2.1 shows how the data for these three factors are organized and analyzed for the Gauge R&R Experiment to obtain the most significant components of variability.

**Table 2.1 Coordinates variability data for the gauge R&R experiment**

| Days | Subject 1 | | | Subject 2 | | | …… | | | Subject m | | |
|---|---|---|---|---|---|---|---|---|---|---|---|---|
| | Test 1 | Test 2 | Test 3 | Test 1 | Test 2 | Test 3 | Test 1 | Test 2 | Test 3 | Test 1 | Test 2 | Test 3 |
| 1 | | | | | | | | | | | | |
| 2 | | | | | | | | | | | | |
| 3 | | | | | | | | | | | | |
| . | | | | | | | | | | | | |
| . | | | | | | | | | | | | |
| . | | | | | | | | | | | | |
| . | | | | | | | | | | | | |
| n | | | | | | | | | | | | |

The design of experiment includes testing of twelve young, injury free subjects. These subjects were selected based on their diverse anthropometric dimensions. These subjects were a range of males and females aging 18-21 years of age. In this study, six females and six males participated in the tests. The subjects were asked to do a 10-foot walk test during three nonconsecutive days. Three walking test trials accomplished for each day with a short break between trials. The raw



data containing 25 joints locations were acquired from a depth camera located in front and three feet away from the end point of their walking path (see Figure 3.3). Once joint data was captured using Microsoft sensors, sample entropy (to be discussed in Section 3.2.5) was then used to summarize the variations into one value to make it easier for comparisons. As an example, the result obtained for Mid Spine (MS) is shown in Tables 2.2, 2.3. The results for the other joints are very similar. The computations were performed using the balanced ANOVA routine in Minitab 2017.

**Table 2.2 Coordinates variability versus days, subjects**

| Two-way ANOVA Table with Interaction | | | | | | Two-way ANOVA Table without Interaction | | | | | |
|---|---|---|---|---|---|---|---|---|---|---|---|
| Source | DF | SS | MS | F | P | Source | DF | SS | MS | F | P |
| Day | 2 | 0.00005 | 0.000027 | 0.087 | 0.917 | Day | 2 | 0.00005 | 0.000027 | 0.119 | 0.888 |
| Subjects | 11 | 1.80098 | 0.163726 | 530.939 | 0.000 | Subjects | 11 | 1.80098 | 0.163726 | 723.608 | 0.000 |
| Day * Subjects | 22 | 0.00678 | 0.000308 | 1.533 | 0.090 | Repeatability | 94 | 0.02127 | 0.000226 | | |
| Repeatability | 72 | 0.01448 | 0.000201 | | | Total | 107 | 1.82231 | | | |
| Total | 107 | 1.82231 | | | | ($\alpha$ to remove interaction term = 0.05) | | | | | |

**Table 2.3 The repeatability and the reproducibility contributions of the gauge**

| | | % Contribution |
|---|---|---|
| Source | VarComp | (of VarComp) |
| Total Gage R&R | 0.0183929 | 100.00 |
| Repeatability | 0.0002263 | 1.23 |
| Reproducibility | 0.0181666 | 98.77 |
| Subjects | 0.0181666 | 98.77 |
| Days | 0.0000112 | 0.00 |
| Total Variation | 0.0184041 | 100.00 |



Table 2.2 shows the two-factor analysis of variance for this experiment, once considering the possibility that the factors (Days and Subjects) may affect the response variable (SE values) jointly, and once considering no interactions between the factors. Since the interaction between "Days" and "Subjects" is not significant at 95% significance level, we use the no interaction model shown in Table 2.2.

Based on the P-values, we conclude that the effect of subjects is large, trials have a small effect, and there is almost no significant effect on variability caused by different days. Table 2.3 reveals the values of two components of measurement error (Daryabari et al., 2019), usually called the repeatability and the reproducibility of the gauge for this experiment. As seen in Table 2.3, reproducibility is the variability due to different subjects using the gauge in this experiment contains more than 99.9% of all variations. In contrast, repeatability which basically indicates the variations due to different trials/days and so reflects the inherent precision of the gauge itself, contains only about 0.1% of the possible variations. The in-depth cameras used in this research are indeed very consistent and reliable for the gait change experiments to be performed in this thesis.

### 2.3.2. Software Drawbacks

Like all systems, the expanded software used in this study has some disadvantages that mostly stem from the camera features. These drawbacks, however, have scarcely been documented. These disadvantages include privacy issues (Kim, 2012), segment length variations and estimation of angles (Bonnechère et al., 2012). However, our exploratory case study investigated some more



possible drawbacks of depth cameras perceived by our team and proposed ways to mitigate those problems in the next section.

It is important to ensure that the depth map is a meaningful 3D representation of the scene (Kadambi et al., 2014). However, sometimes the coordinates do not seem to be meaningful, particularly at the beginning and the end of each sample test when the coordinate data suddenly jump. This is the biggest issue we faced with, in the observations collected. We used a simple filtering strategy to trim the data. We also had restrictions based on lighting and camera angles. If the room where we conduct tests at is not well-lighted, then we might observe non-accurate coordinate data as the system is not able to track the exact position of the joints, so it will just infer the approximate locations based on the walking trends. We tested in a well-lit room to make sure enough definition is available to the camera. As well as keeping the camera level at a set height to account for the errors that could span from a changing camera angle. There are some other drawbacks that are mentioned in the following of the current study. These limitations must be accounted for before we can expand the usage of the proposed system to other environments and situations.

The literature survey and the capability study conducted in Chapters 1 and 2 demonstrate that:

1. Analyzing people's gait provides essential information for measuring both mental and physical health as well as the progress of physical therapy and rehabilitation,

2. Many researchers have used 3D kinematic data obtained by Microsoft Kinect devices in measuring activities and postural analysis studies,



3. Microsoft Kinect<sup>TM</sup> devices may provide a reliable way to capture walking patterns of a person, and

4. Microsoft Kinect<sup>TM</sup> is an accurate and affordable tool for the assessment of human gait.

Thus, in this thesis, we used Microsoft Kinect<sup>TM</sup> to analyze a group of people's gaits, detect and quantify possible changes, and identify different persons.



# Chapter 3.  Experimental Plan and Proposed Methodology

This chapter depicts how to set this system up and use it to collect the required data, clarifies what data to collect, and briefly explains how to process the collected data into statistics suitable for gait analyses.

## 3.1. Experimental Plan

Two Microsoft Kinect cameras were used to create the 3D Kinematic view of 25 joints from a human body. The depth cameras and other related devices were placed in an appropriate location in the Ergonomics Lab at Kansas State University. This protocol remained the same for all the conducted tests. However, the group of subjects were different from those who participated in the previous experiment explained in Chapter 2. Nevertheless, some subjects remained the same.

In this study, participants completed a series of testing sessions by walking back and forth normally, through a 10-foot path one time with the help of a motion-restricting device or walking aid and another time without any device. They were asked to perform each test a few times and instructed to wear fitted clothing and athletic shoes. The data collection sheet is given in Table 3.1. The assessments, devices, participants, protocol, setup process and data analysis are described in detail in the following sections.

### 3.1.1. Participants

Ten healthy students consisting of five females and five males as volunteers performed a 10-foot walk test. The average height, shoe size, and age and their corresponding standard deviation are 5 8" (2.3") feet, size 8.7 (1.2) and 21 (1.6) years old, respectively. No subject had any history of



medication use, neurological conditions, musculoskeletal disorders, or major injury prior or during the experiments.

### 3.1.2. Motion-Restricting Devices

For this study, an ankle brace, an ACL brace, an adjustable (dynamic) hinged brace, a walking cane and a rolling walker were used as the motion-restricting devices. Figure 3.1 illustrates an ACL brace on the left and an ankle brace on the right side. Also, Figure 3.2 illustrates a cane in the left and a rolling walker used in this study on the right side.

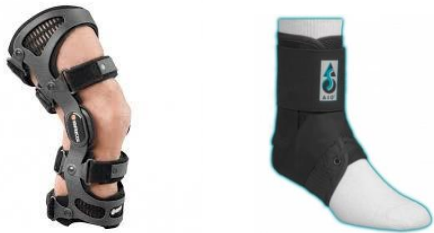

**Figure 3.1 Motion-Restricting Devices (ACL Brace, and Ankle Brace)**

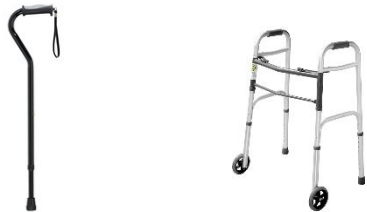

**Figure 3.2 Walking Aids (Cane and Rolling Walker)**

### 3.1.3. Setup Processing

In this study, a simple walk test was conducted on ten healthy subjects. The layout of the testing area is shown in Figure 3.3, in which two camera angles are used to record the subjects' walking data. The testing area includes a 10-foot path that each subject walks through, turns around, and



walks back. As seen in Figure 3.3, one Kinect camera is placed eight feet away from the walking path to the side and the other one is placed in front and three feet away from the endpoint.

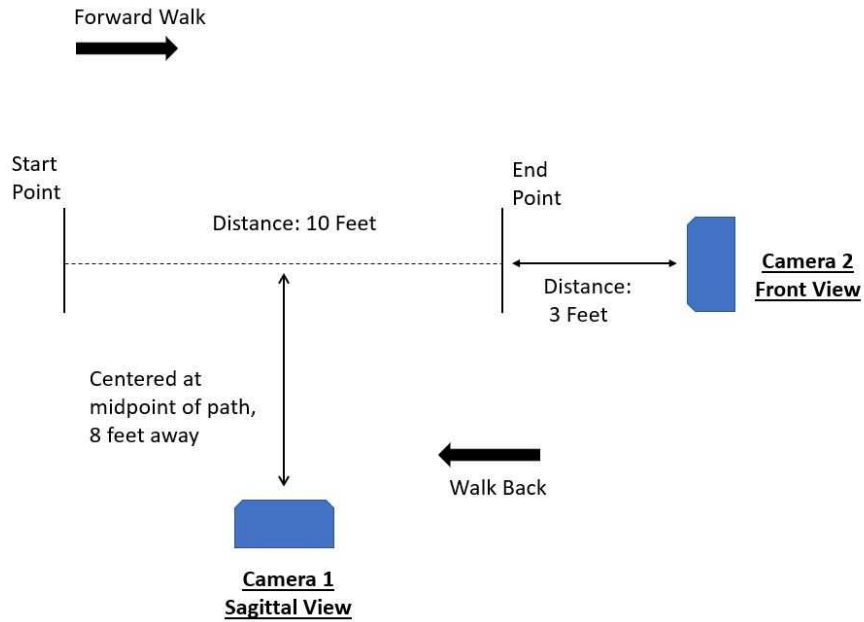

**Figure 3.3 The experimental layout**

All ten subjects were instructed to wear tennis shoes, fit shorts, and a T-shirt to ensure that the system can capture the joints accurately. The subjects were also asked to repeatedly wear similar clothing that would not bias the results of the experiment. The test subjects were instructed when to begin the test and a physical marker was placed near the end of the test area, so the subjects were aware of where they needed to stop without having to look down, which we found to skew results. The test subjects were also asked to lead with the same foot for each walk. They were instructed to walk at a consistent pace. Operators of Kinects counted down from three to ensure consistent recording of test subjects.



## 3.2. Proposed Methodology

### 3.2.1. Research Questions

The first question behind this study is whether Kinect devices can create a personalized walking profile by recording and tracking skeleton joint position data. This entails accuracy, precision, reliability, and ease of use of the system. All these factors are being considered to determine if this system is a practical choice for gait analysis. Through the literature survey and our own experiments (Malmir and Chang, 2019a), we can confirm that Microsoft Kinect$^{TM}$ is capable of generating consistent results under the same circumstances in terms of the proposed sample entropy method. Another question is whether this profile based on individual joints can be used in quantifying changes in human gait instead of gait parameters that were widely used in many studies (Maki, 1997; Cham and Redfern, 2002; Plotnik et al., 2007; Shull et al., 2014). The last question is whether the proposed procedure is capable of differentiating different persons through multiple joints.

### 3.2.2. Data Type

Among 25 joints collected by Kinect, the coordinates of fifteen main relevant joints consisting of head, neck, left shoulder, right shoulder, shoulder spine, mid spine, base spine, left hip, right hip, left foot, right foot, left knee, right knee, left ankle, and right ankle were tracked over time to create a profile of the human body. We did not consider hand-related joints such as wrists and elbows as they are not much involved in a walking practice. Moreover, there is no need to index time as Microsoft Kinect$^{TM}$ was already set to record 30 fps of all selected joints in both directions, so the number of points gathered will show how long it takes for someone to walk the 10-foot testing path.



Among three possible axes, only X (anteroposterior) and Y (vertical) dimensions are considered for further statistical analyses for different purposes throughout this study. Examining the variability in data measure due to the displacement of joints, we have found that variations in Y dimension are far less than those in X and Z (mediolateral) dimensions. This is the main reason why Y is chosen as the primary indicator of joint variations. Table 3.1 shows the type of data (Raw data or gait parameters) used for different analyses conducted in the next chapter.

**Table 3.1 Required data and axis for different studies**

| Study | Data Type | Required Axes | Camera Angle |
|---|---|---|---|
| 1. Gait Changes Detection | Raw Data | Y | Sagittal |
| | Gait Parameter | X and Y | Frontal |
| 2. Gait Recognition | Raw Data | Y | Sagittal |
| | Gait Parameter | X and Y | Frontal |

Figure 3.4 shows three main perspectives of joints locations obtained from different camera angles.

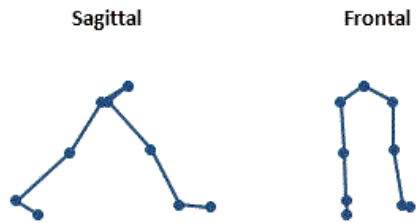

**Figure 3.4 Camera angles**



### 3.2.3. Skeleton Profiles

Evidently recording the coordinates of a joint over time makes a time series. Figure 3.5 depicts eight selected joints in Y coordinate of a female subject walking ten feet. It demonstrates the vertical positions of this subject during the walk.

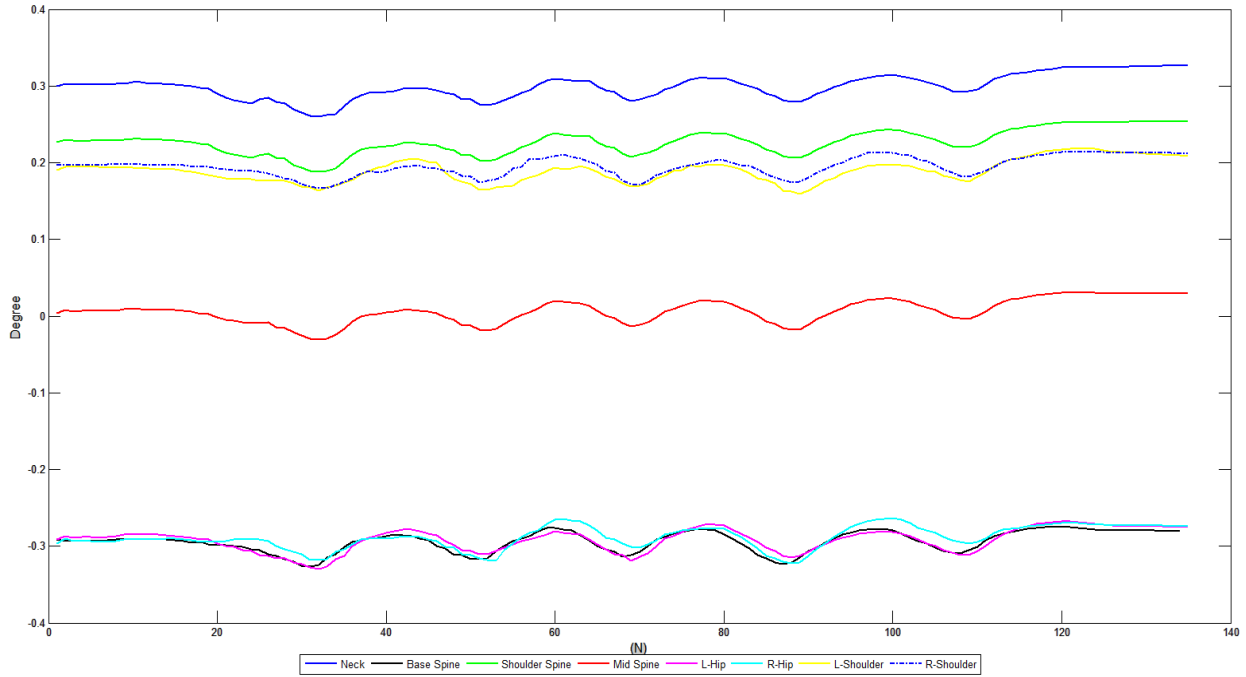

**Figure 3.5 Tracking eight selected Kinect joints of a healthy female subject in Y coordinate over time**

Gait is a key factor in determining the overall health of a subject (Zeni & Higginson, 2009). Therefore, the creation of a personal gait profile would be helpful in tracking personal well-being, particularly for the elderly population. For example, large changes in an elderly person's gait profile may be an indication of elevated fall risk. These profiles may also be used to track progress in a series of physical therapy sessions.



### 3.2.4. Gait Parameters

As discussed earlier, Microsoft Kinect$^{TM}$ was used to track fifteen main joints over time in this study. We use the coordinates of these joints to make some gait parameters that are proposed to detect gait changes in the next chapter. Although these joints formed a human skeleton, only some of them were relevant to make selected gait parameters. Table 3.2 lists these gait parameters and the relevant joints generated by Kinect with corresponding gait parameters through the frontal plane (Chattopadhyay et al., 2015).

**Table 3.2 Descriptions of gait parameters**

| Gait Parameter | Assigned as | Relevant Joints |
|---|---|---|
| Spine Tilt | $V_1$ | Shoulder Spine (ShS) - Base Spine (BS) |
| Hip Tilt | $V_2$ | Left and Right Hips (LH-RH) |
| Shoulder Tilt | $V_3$ | Left and Right Shoulders (LSh-RSh) |

X and Y coordinates generated from Kinect joints ShS, BS, LH, RH, LSh, and RSh listed in Table 3.2 can be converted into gait parameters such as spine tilt, hip tilt, and shoulder tilt. The X dimension typically tracks the axis perpendicular to the camera's line of sight, whereas the Y dimension tracks the up-and-down movement of a person, no matter where the camera is placed at. Y and Z coordinates of the joints may also be considered in cases where Kinect is used from the Sagittal perspective i.e. Camera 1 shown in Figure 3.3. Z-direction tracks if a subject is closing in or fleeing the camera. In this study, however, only X and Y directions were necessary.



### 3.2.4.1. Gait Parameter Processing

In anatomy, the hip tilt is the orientation of the pelvis in respect to the femurs it rests upon in space (Malmir and Chang, 2019b). Hip tilt, spine tilt, and shoulder tilt within 5 degrees each are typical for walking, meaning that the respective tilts should be no more than 5 degrees for a healthy subject (Abdi, 2002). Deviation over 5 degrees indicates that a subject may suffer a physical problem in walking properly.

As mentioned, all subjects in this study were healthy. In fact, one of the study goals was to analyze the measured tilt of their spines, hips, and shoulders during walking and compare those measurements to other healthy subjects. In addition, differences in tilt measurements on the same subject with and without a knee brace on his right knee are analyzed in Chapter 4. Gait parameters and relevant joints are illustrated in Figure 3.5.

The mathematical relation between relevant joints in order to acquire gait parameters in the frontal plane is

$$y_2 - y_1 = m(x_2 - x_1),$$

(3.1)

where $(x_i, y_j)$, $I = 1,2$ is the coordination of two selected Kinect joints that were considered to be the connection vectors and were tracked over time. Specifically, $x_1$ and $x_2$ were assigned as the X coordinates of the left- and right-sided joints, and $y_1$ and $y_2$ were assigned as the Y coordinates of the left- and right-sided joints, respectively. Parameter $m$ is the slope of these vectors, changing over time. Then $m$ was converted to the angle between two joints at each point by the relation $m = \tan\alpha \Leftrightarrow \alpha = Arctan\, m$,

(3.2)



where $\alpha$ is a radian-based measure. In order to convert $\alpha$ to a degree, we multiplied it by $180/\pi$.

MATLAB (R2014a) was used for computation. Note that equation (3.1) is applied to each row in the data table related to each frame of skeleton. Thirty sets of angles are generated for each second of walking.

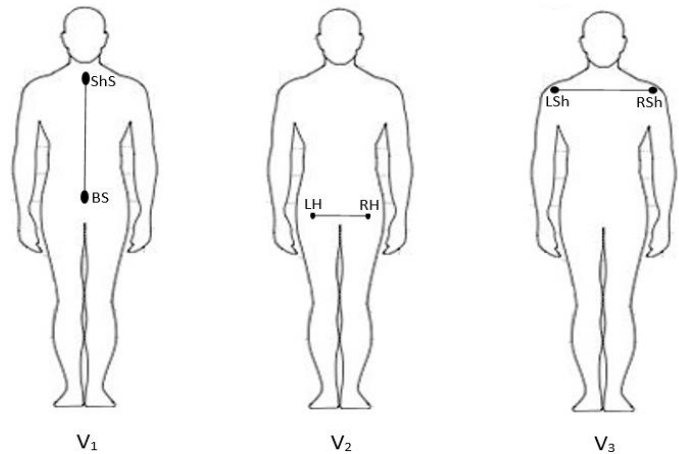

**Figure 3.6 Relevant joints of proposed gait parameters**

### 3.2.5. Sample Entropy

The joint angles in time series are then converted into a statistic that measures the variability or changes. We proposed to use sample entropy (SE) as an estimator of this statistic. The proposed SE is capable of measuring the deterministic or stochastic content of a time series (regularity), as well as the degree of structural richness (complexity), through operations at multiple data scales (Looney et al., 2018).

Following the limitations of standard entropy measures for short and noisy time series, the approximate entropy method was presented by Pincus (1991). It specifies the probability that in a time series, analogous patterns and the signal delay vectors (DVs) will stay analogous when the



pattern lengths are raised. A strong expansion, which disregards self-matches, named the SE, has been developed (Richman and Moorman, 2000) and is defined as follows:

1. For lag t and embedding dimension m, generate DVs:

$$X_m(i) = [x_i, x_{i+1}, \ldots, x_{i+\tau(m-1)}]$$

where $i = 1, 2, \ldots, (N-\tau(m-1))$.

2. For a given DV, $Xm(i)$, and a threshold, r, count the number of instances, $\phi m(i, r)$, for which $d\{Xm(i), Xm(j)\} \leq r, \ i \neq j$, where $d\{0\}$ denotes the maximum norm.

3. Define the frequency of occurrence as

$$\phi_m(r) = \frac{1}{N - \tau(m-1) + 1} \sum_{i=1}^{N - \tau(m-1) + 1} \phi_m(i, r) \tag{3.3}$$

4. Extend the embedding dimension $(m \rightarrow m + 1)$ of the DVs in step (1), and repeat steps (2) and (3) to obtain $\phi_{m+1}(r)$.

5. The SE is defined as the negative logarithm of the values for different embedding dimensions, that is,

$$SE(m, r, \tau) = -\ln\left[\frac{\phi_{m+1}(r)}{\phi_m(r)}\right] \tag{3.4}$$

In general, the less predictable or the more irregular a time series, the higher its SE. A block diagram of the algorithm is shown in Figure 3.7.



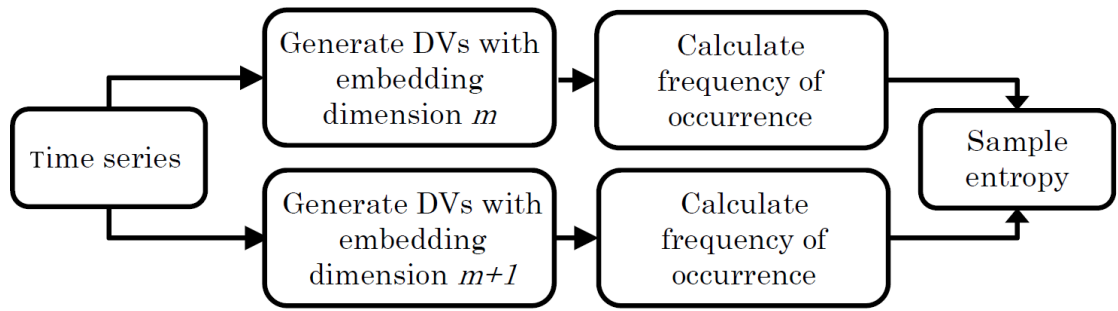

**Figure 3.7 Block diagram of the sample entropy algorithm (Looney et al., 2018)**

In the current study, sample entropy converts each series of recorded displacements for a joint over time into only one value which makes them comparable together. This potential of SE besides calculating variations within the joints (Ramdani et al., 2009) and variations due to the displacement has made it a very decent algorithm for comparing different gait conditions, quantifying changes, and identifying different subjects.



# Chapter 4.  Studies of Gait Change Detection and Identification of Distinct Persons through Gaits

The proposed gait measurement methods in sections 3.2.4 and 3.2.5 quantify variabilities due to the displacement of a joint during a walk. These values are used to compare variations between subjects and within a subject obtained through several tests implemented in different days. The variabilities within a person help us track their physical health and can assist with advancing rehabilitation while the differences between subjects help to classify individuals.

Hence, the quantification process is essential for tracking joint health, especially for individuals who are undergoing physical therapy and are affected by an age-related disability. This quantification strategy can be extended to other physical health-related areas. It can help monitor the elevated risk of falling as reflected in gait changes due to physical weakness. For example, to examine gait changes in persons with multiple sclerosis (MS) who have minimal disabilities. Sosnoff et al. (2012) have already found out that persons with MS walked with fewer, shorter, and took wider steps and had a greater variability in the time between steps than healthy individuals. Those characteristics may provide much more identifiable differences in walking patterns than age and gender.

This chapter covers three main studies conducted in the current thesis. The first study explores three analytical approaches for detecting differences in gait changes. Both gait parameters and sample entropy methods were used to quantify human walking gait changes. The goal is to determine if gaits with and without motion-restricted devices are different.



In the third study, we explored several classification models using supervised learning to categorize and predict if gaits belong to the proper person. The goal is to determine if gaits can be used to distinguish two persons of the same physical profile. The second goal is to determine if the supervised learning models can be used to distinguish a group of people.

## 4.1. Study 1: Gait Change Detection using Parameters Obtained from Depth Cameras

Study 1 investigates whether the proposed method can detect a person's gait change, through two analytical approaches using either gait parameters or sample entropy measures. First, we proposed to use the gait parameters as mentioned in section 3.2.4, the data derived from Kinect for each joint makes a profile containing multiple joints. This profile is transformed into numerical values for comparisons. To this aim, three different experimental tests were conducted using depth cameras on the same subjects in Study 1.

### 4.1.1. Data Collection

For the first analytic experiment, a 10-foot walking test was implemented on ten healthy subjects introduced in Section 3.1.1, to detect possible changes in their walking patterns with and without motion-restricted devices. The subjects walked forward through the 10-foot path one time wearing an ACL brace on their right knee (first test) and another time without the brace (second test). This procedure was replicated three times. For the second experiment, the same subjects performed the 10-foot walking test three times, using all motion-restricting devices introduced in Section 3.1.2, and once without any devices.



## 4.1.2 Analytic Method 1: Gait Change Detection using Gait Parameters

Skeleton coordinates of all joints of a subject were generated by the Kinect skeleton feature, and the three gait parameters, spline tilt, hip tilt, and shoulder tilt (Table 3.2), were calculated for each recorded skeleton frame. Then, SE measure on the time series of each gait parameter was obtained. Final results of SE measures based on gait parameter are shown in Figure 4.1-4.3, where $T_{ij}$ indicates the distribution of the three trials for each experimental setting according to

$$T_{ij} \begin{cases} Subjects & i \in (0,9) \\ Test\ No & j \in (1,2) \end{cases}$$

where the indices $i$ and $j$ indicate the subjects' IDs and the test number, respectively. Test number 1 represents a subject wearing an ACL brace while test number 2 represents normal walking without wearing the device.

Figure 4.1-4.3 show prominent differences between SE values of spine tilt, hip tilt, and shoulder tilt for all subjects in two different conditions. The variability range of SE values in the first test (T1) was larger than the variability range of SE values in the second test (T2) for all subjects. Therefore, any gait parameter introduced in Section 3.2.4 may be a potential candidate to measure physical therapy progress. However, combinational use of all three gait parameters on star glyphs may provide more distinction power of gait changes when the changes are small.



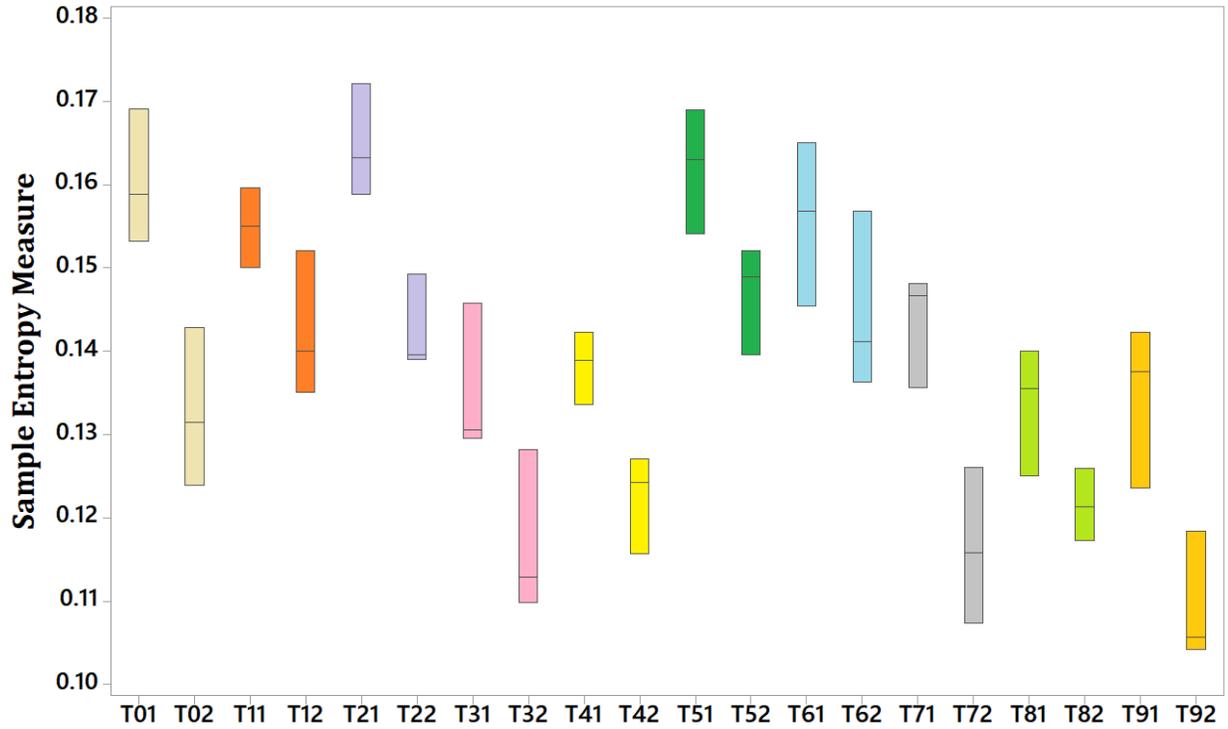

**Figure 4.1 Comparison of the subjects' walking status in two different conditions using SE measures of <u>spine tilt</u>**

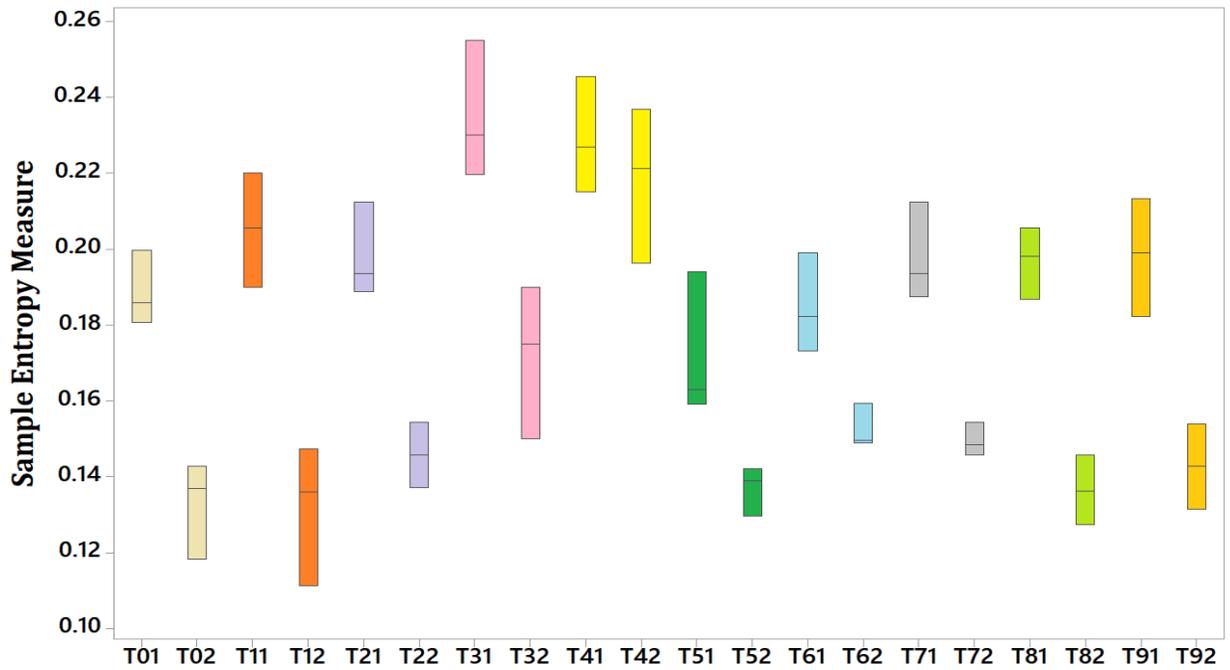

**Figure 4.2 Comparison of the subjects' walking status in two different conditions using SE measures of <u>hip tilt</u>**



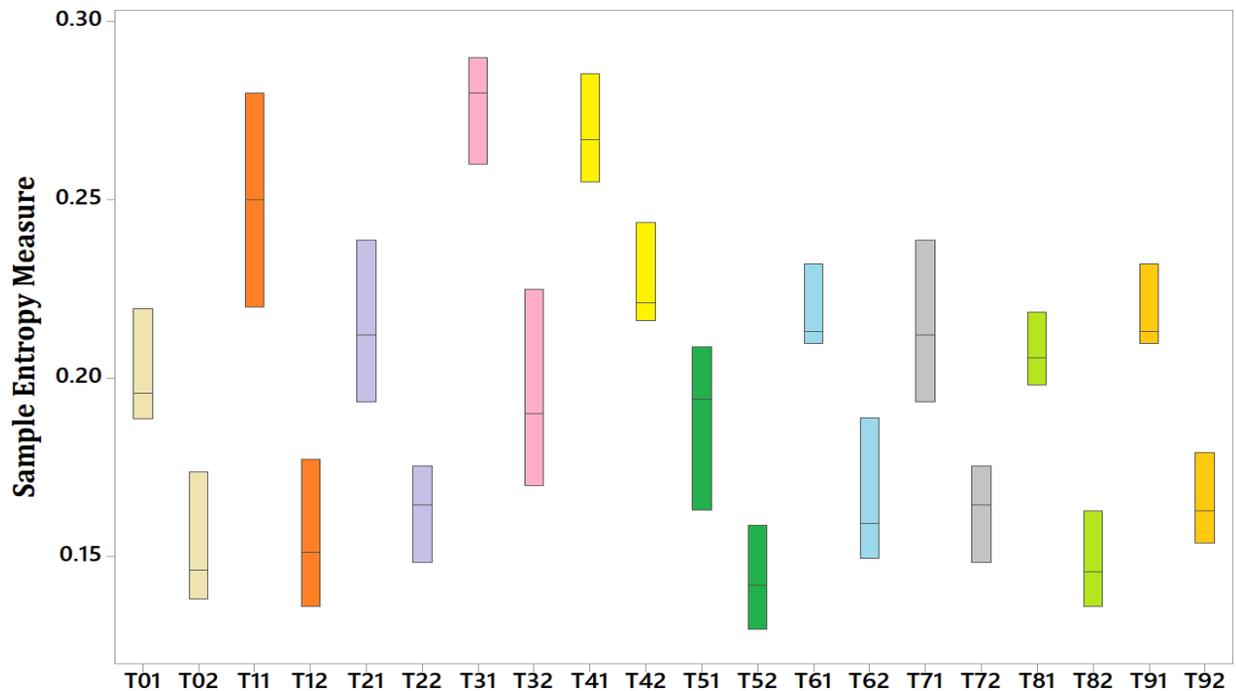

**Figure 4.3 Comparison of the subjects' walking status in two different conditions using SE measures of <u>shoulder tilt</u>**

The Hotelling $T^2$ statistic (Montgomery, 2016) could be used for a retrospective analysis of the mean vector of SE measures of three gait parameters in normal walking condition as the phase 1 analysis. This practice establishes an "in-control" measure for normal walking. Then we could consider the ACL brace-related walking data as the "out-of-control" condition for abnormal walking in an attempt to detect changes simulated by wearing the ACL brace. However, three replicates obtained from ten subjects does not contain enough observations for these analyses. Therefore, adding more replicates for each subject may provide a more accurate in-control statistic to identify possible changes in gaits. Each subject should then have his/her own Hotelling $T^2$ or other multivariate control charts (Montgomery, 2016).



### 4.1.3 Analytic Method 2: Gait Change Detection using SE on Raw Coordinates Data Derived from Microsoft Kinect™

An analytic method based on the SE on raw Kinect data was applied to the same ten subjects who performed the 10-foot walking test three times. However, in addition to the knee brace, other motion-restricting devices introduced in Section 3.1.2 such as the ankle brace, cane, and walker were also included. Moreover, sample Entropy was applied to the raw data derived from Microsoft Kinect™ directly, instead of the gait parameters. This data contains the coordinates of 25 joints for each subject at each frame over time. Note that we have expanded the joints from 15 to 25 in this study.

In this experiment, to see if the system can capture the changes due to using motion-restricting devices, the following hypothesis was considered

$$H_0: D_{ik}=0 \text{ vs } H_1: D_{ik} \neq 0$$

$$D_{ik} = \sum_{j=1}^{25} NW_{ij} - \sum_{j=1}^{25} MD_{ijk} \qquad (4.1)$$

where k indicates an ID for each type of motion-restricting device (k $\in$ {AB, KB, Walker, Cane}), i is the index for subject (i=1, 2, …, 10 as opposed to 0 to 9 used in the previous section), and j=1,2, …, 25 is the index for the joints. Thus, equation (4.1) tests whether the sum of SE measures of normal walking (NW) is different from that of motion-restricting walking (MD). Note that the more joints are included the more aggregated differences are expected to be generated.

Figure 4.4-4.7 show 95% confidence intervals of the D statistic for each subject notated by $S_i$, i=1,2,…,10.  If an interval does not cover zero, it indicates significant differences between the



normal walking (NW) and motion-restricting walking (MD). For the most parts, the proposed method is capable of separating NW and MD except for subject 10 wearing ankle brace, S3, S7, and S9 wearing a knee brace. In these cases, the summary statistic D shown in equation (4.1) may not be able to distinguish normal gaits from motion-restricting gaits. Instead, individual SE values on different joints plotted on star glyphs may be used to further separate the walking patterns. Details can be found in the next section.

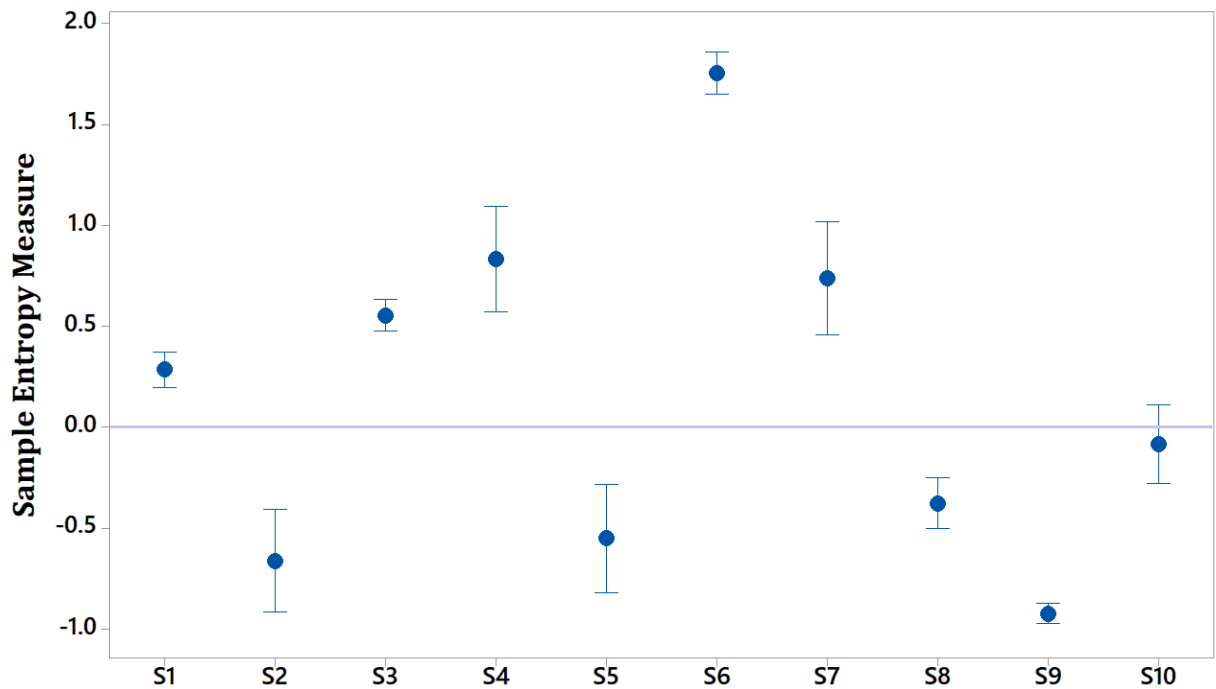

**Figure 4.4 Deviations of a set of normal walking trials from a set of abnormal walking trials caused by using an <u>ankle brace</u> (AB) on the subject's right knee**



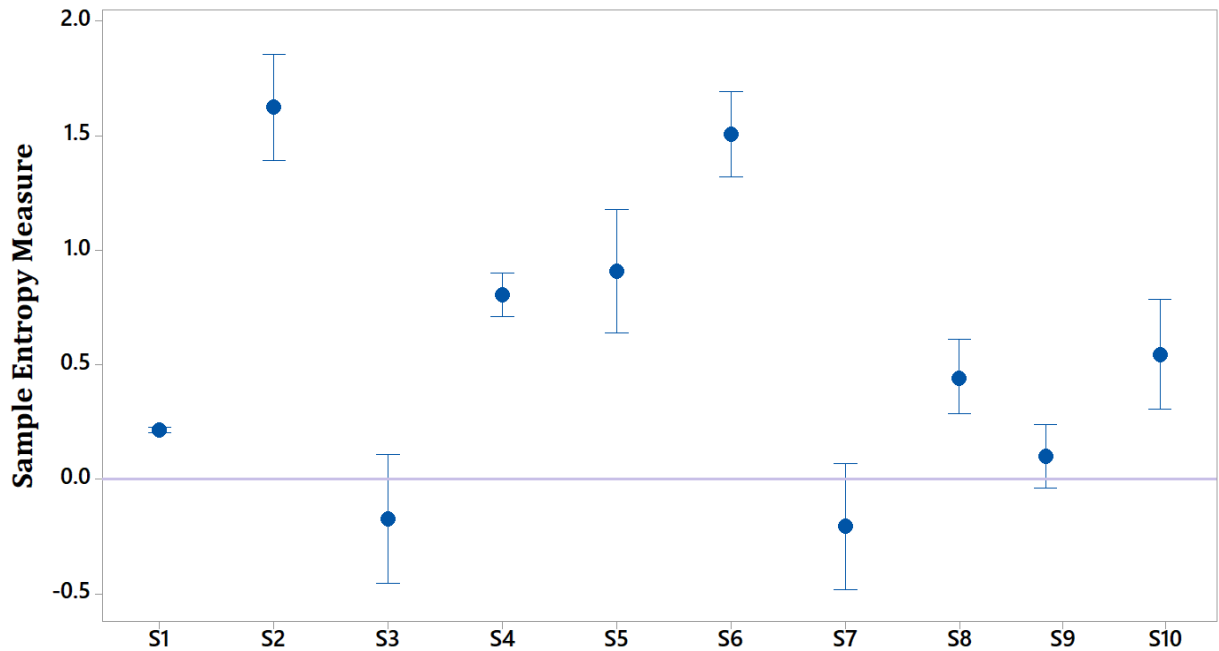

**Figure 4.5 Deviations of a set of normal walking trials from a set of abnormal walking trials caused by using an <u>ACL brace</u> on the subject's <u>right knees (KB)</u>**

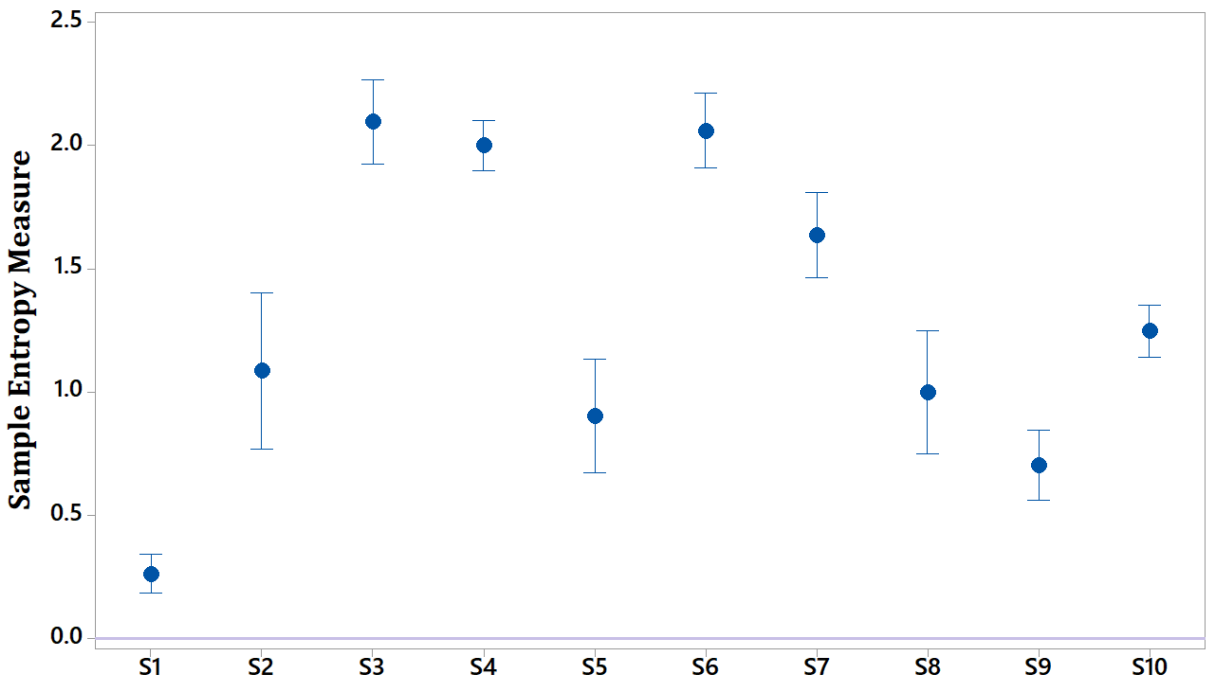

**Figure 4.6 Deviations of a set of normal walking trials from a set of abnormal walking trials caused by using a four-legged walker (Walker)**



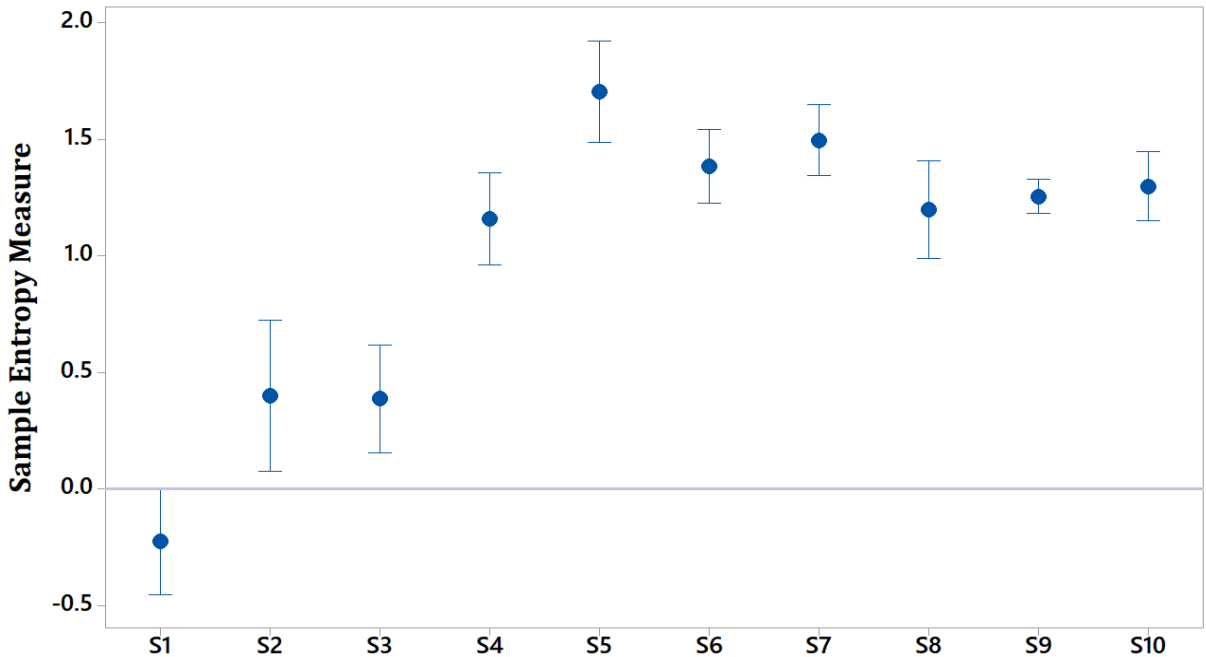

**Figure 4.7 Deviations of a set of normal walking trials from a set of abnormal walking trials caused by using a <u>walking cane (Cane)</u>**

### 4.1.4 Analytic Method 3: Finding the Significant Joints to Identify Changes in People's Walking Patterns

In previous analyses in sections 4.1.2 and 4.1.3, summary statistics were used to distinguish whether gaits of normal walking change when a motion-restricting device is applied. The next logical question is that which joint or joints would contribute to the difference? To answer this question, the coordinate data associated with three male subjects out of the group of subjects tested in the previous experiments were randomly selected to be compared in three different conditions. One condition is the normal walking and the rest are walking with the help of the ankle brace and the ACL brace introduced in Section 3.1.2. Also, fifteen main joints were considered as the significant ones with higher impacts on discriminating a normal walking from an abnormal walking, especially for walks with an ACL or ankle brace.



Subjects 1, 2, and 3 denoted as *S1*, *S2*, and *S3* are shown in Figure 4.8 (A), (B), and (C), respectively. The notation of subjects wearing ACL brace is denoted as *KB* shown as a solid line. The test for ankle brace is denoted as *AB* shown in dotted lines. Finally, the normal walking test is denoted as *NW* shown in dashed lines.

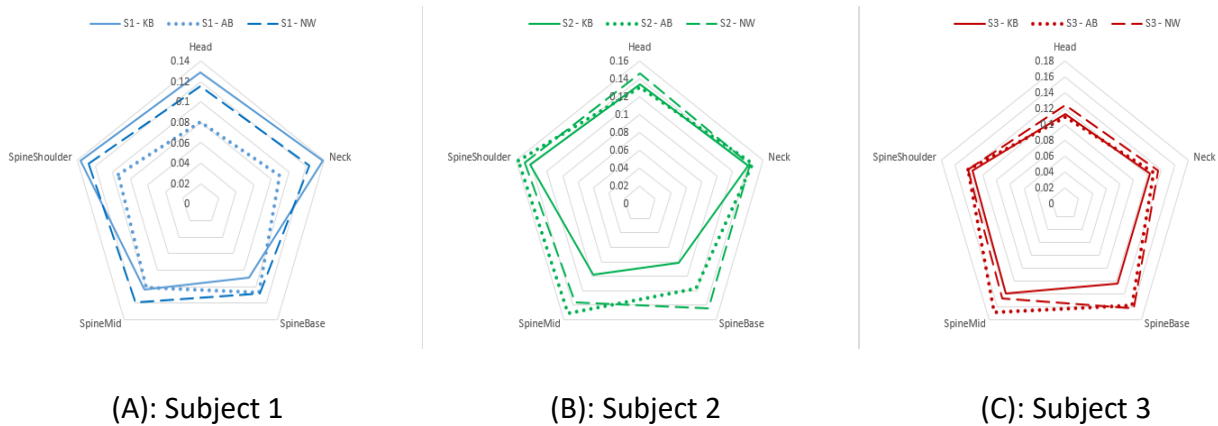

(A): Subject 1          (B): Subject 2          (C): Subject 3

**Figure 4.8 Comparison of the mean SE values of the profiles of five middle joints in three different conditions (NW: Normal Walking; AB: Ankle Brace; KB: Knee Brace)**

Out of the fifteen joints, a group of **upper body joints** consisting of head, neck, base spine, mid spine and shoulder spine was the primary group of joints used to quantify changes with the least amount of variance. Figure 4.8 shows differences between SE values of these five joints for each person under three different conditions. Some SE values are pretty close while some others have significant differences. The variability of SE values in the base spine and mid spine in different conditions were larger than the variability of SE values in the other joints. Therefore, these two joints may be the potential candidates to measure physical therapy progress. The combinational use of all joints on star glyphs may provide a more concise presentation of gait changes if tracked over time. However, since the braces were worn on the right side, **five main joints of the left side**, as well as **five main joints of the right side** of the human body were considered for further



analysis in this analytic task. All the results discussed so far were obtained from the side camera. The same procedure was done for statistical analyses of the data on the vertical dimension derived from the frontal camera as well. The results were similar to those from the side camera in terms of patterns.

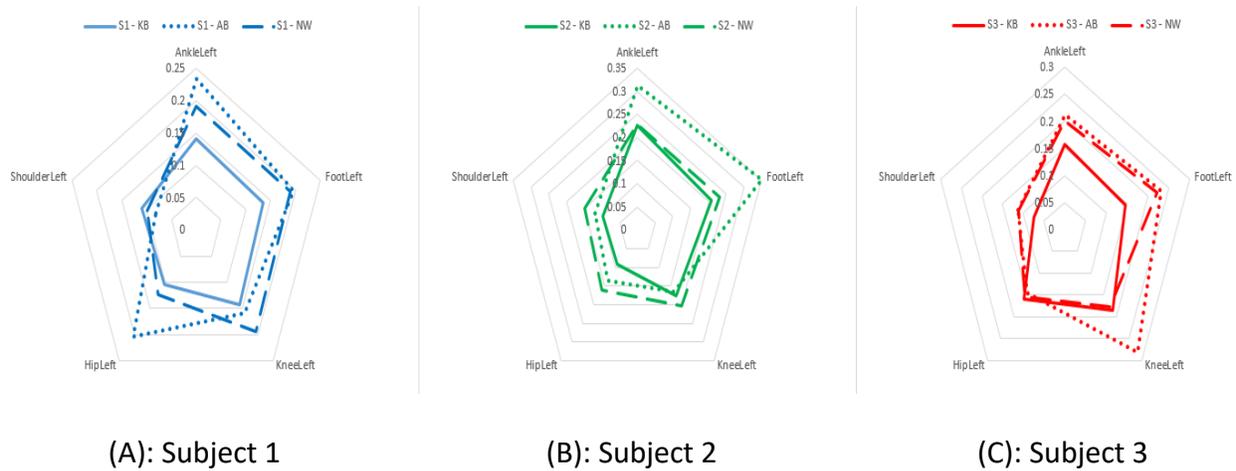

(A): Subject 1          (B): Subject 2          (C): Subject 3

**Figure 4.9 Comparison of the mean SE values of the profiles for five left side joints in three different conditions (NW: Normal Walking; AB: Ankle Brace; KB: Knee Brace)**

Figure 4.9 illustrates the differences between SE values of five left side joints for each person under three different conditions. As seen in the figures, affected joints when each person wears a brace and walks is different in different subjects. These five lower-body joints distinguish normal walking from motion-restricting devices better than those using the upper-body joints.

Figure 4.10 demonstrates the results obtained from the same analysis, but on five right side joints of each individual. Similar to the analysis of the middle body joints, some SE values look pretty close to one another while some others have significant differences. For example, there are almost no differences between the variability of right knee, right hip, and right shoulder of Subject 3 when this subject uses either one of the motion-restricting devices. However, he did



not have a consistent walking pattern as expected since SE values of his right foot and right ankle were different.

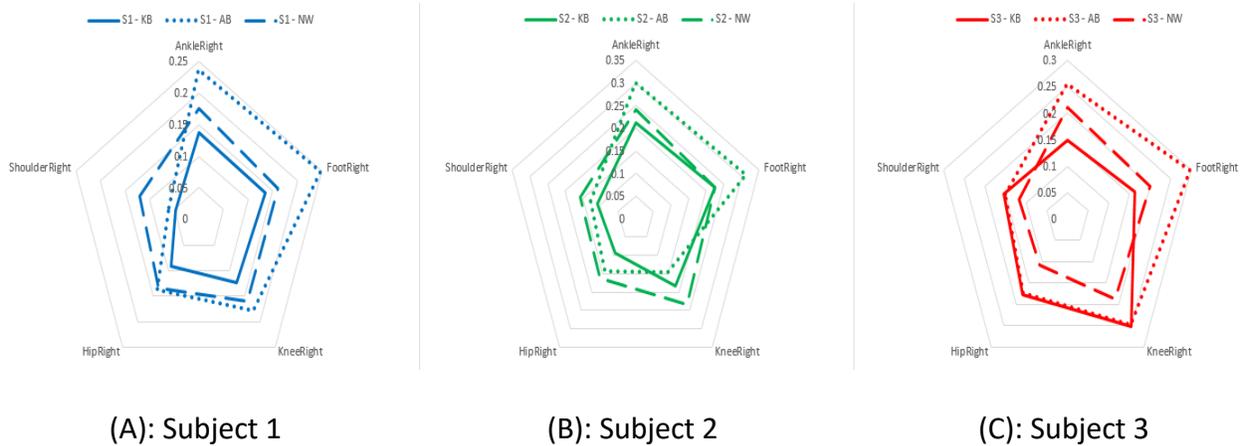

(A): Subject 1          (B): Subject 2          (C): Subject 3

**Figure 4.10 Comparison of the mean SE values of the profiles for five right side joints in three different conditions (NW: Normal Walking; AB: Ankle Brace; KB: Knee Brace)**

As seen in Figures 4.8-4.10, subjects have different walking patterns. Profiles consisted of SE values from various joints show that motion-restricting devices do alter their gait patterns since the pentagon shapes of KB and AB were much different than NW. However, comparing Figures 4.9 and 4.10, we observe a relatively similar pattern in pentagons related to the same five joints in the left and right side of each individual's body in all three different conditions. The same scale is considered for each subject in two cases to provide fair comparisons.

We next compare the lower body joints from the left-hand side and the right-hand side. Equation 4.1 was applied to the differences from normal walking of the five left side joints (shown in Figure 4.9) and those on the five right side joints (shown in Figure 4.10) in two different conditions. The first condition is when the subjects wore an ankle brace (k=1). The differences are shown in Figure



4.11. The other condition is the subjects wore an ACL brace (k=2) while walking. The results are shown in Figure 4.12**.**

$$D_{ik} = \sum_{j=1}^{5} NW_{ij} - \sum_{j=1}^{5} MD_{ijk} \qquad (4.2)$$

Figure 4.11 demonstrates that the ankle brace influences on the left-side joints more than the right-side joints as the magnitude of their total deviations from normal walking is a little more. However, there is no significant statistical difference between the left-hand side and the right-hand side since all 95% confidence intervals overlap with each other when subjects wore an ankle brace. On the other hand, the results are totally different when the subjects wore an ACL brace. Figure 4.12 demonstrates that wearing the ankle brace alters the right-side joints much more than the left-side joints as they are much noisier than the left-side joints compared to the normal walking condition. In this case, S1 and S3 have different patterns in the left-hand and right-hand side.

Note that variability between and within subjects are different in terms of the SE values on the raw Kinect data. In fact, the SE values within a person are far less than those from another person. This fact allows us to properly identify individuals. While looking at the variability within subjects, we focus on entropy changes of various joints, that might contribute the most to the changes. The differences within subjects may be the key to document rehabilitation progress.



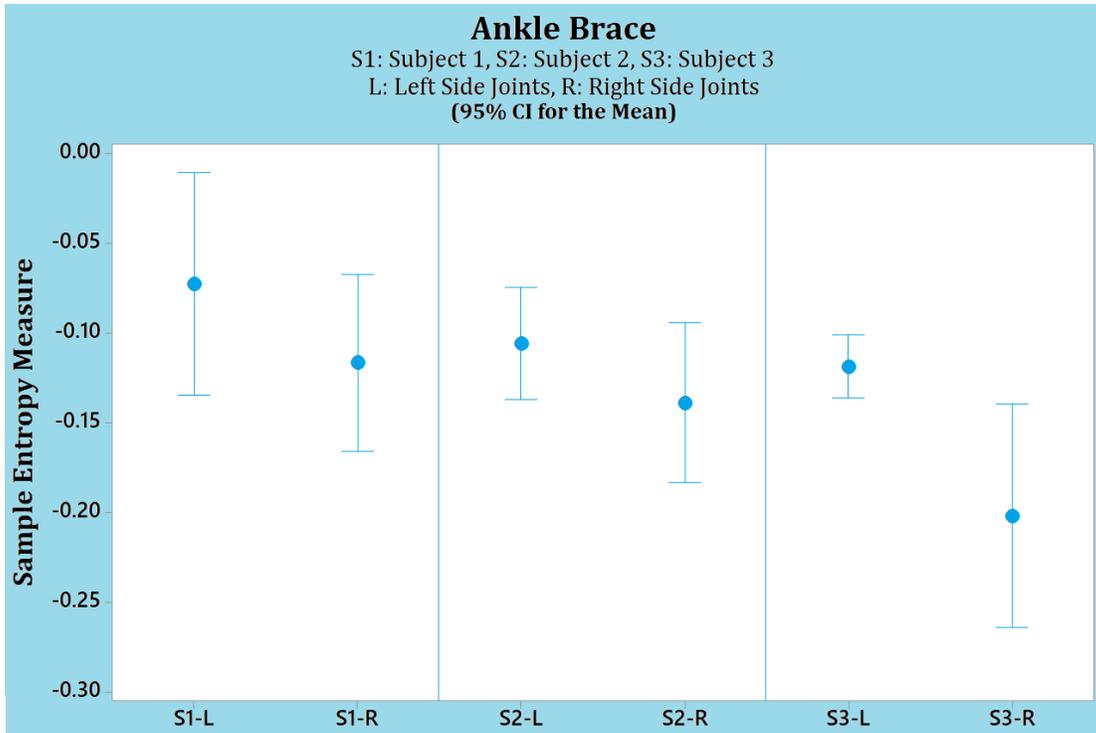

**Figure 4.11 The effects of the motion-restricting devices (The ankle brace) on the right and left side joints**

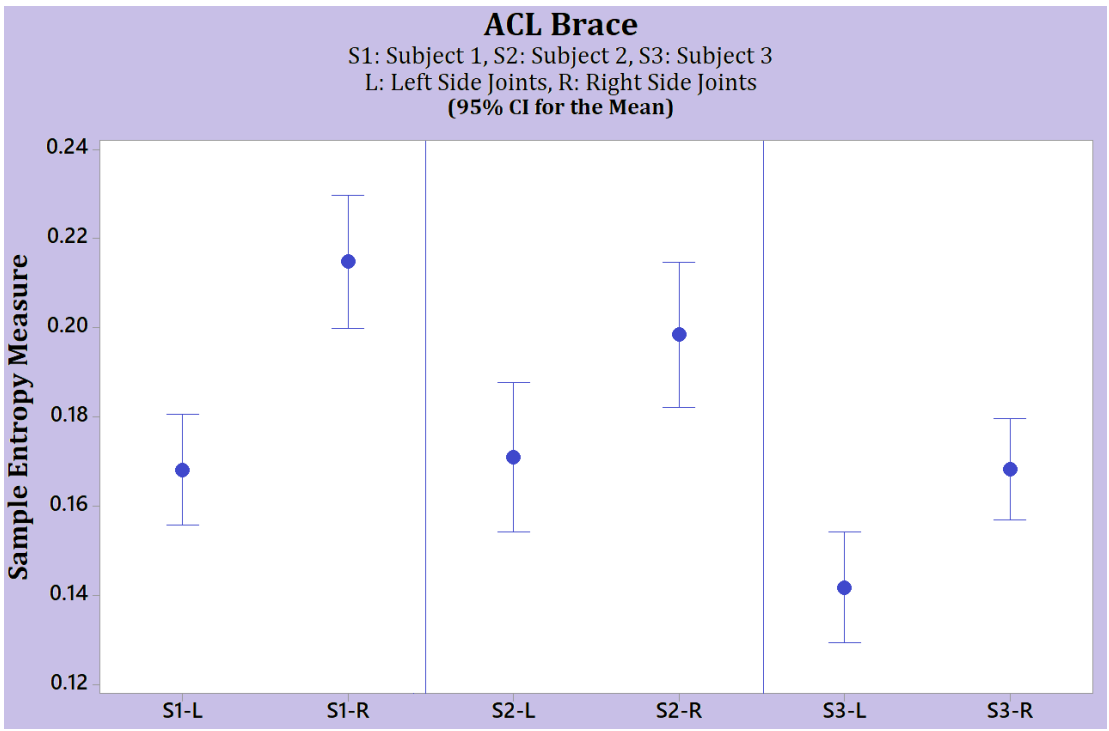

**Figure 4.12 The effects of the motion-restricting devices (The ACL brace) on the right and left side joints**



## 4.2. Study 2: Identifying Different Persons using the Gait Parameters

The goal of this study was to differentiate two subjects' movements based on their gaits rather than physical profiles. Data were derived from Microsoft Kinect™ coordinates by recording 25 joints of the human body over time. Once Kinect™ coordinates were converted into gait parameters considering their relevant joints, spline, hip, and shoulder tilts (i.e., the slope between two relevant joints) were measured over time. Then Sample Entropy (SE) measures were computed based on the angle or tilt changes over time (Aboy et al., 2007; Richman and Moorman, 2000). In this study, these SE measures the spine, hip, and shoulder tilt changes were used to distinct gaits of two persons with similar physical traits.

In order to demonstrate the operation of the proposed method, this study was conducted in two main steps. First, walking patterns were recorded from two subjects via the proposed Kinect™ device, as described in Section 3.2, and the data were used to determine whether or not their walking patterns differed. The subjects chosen were completely healthy with similar physical conditions such as age, gender, height, and weight. Both subjects were instructed to wear tennis shoes and walk in a straight 10-foot path. A Microsoft Kinect™ camera was placed in the frontal direction. Each experimental setting for a gait parameter was repeated 10 times.

Although regression models may be useful for finding significant relationships between gait parameters and factors such as age, gender, height, and weight, they cannot be used to identify differences in their walking patterns. The experimental results showed promise of using gait parameters over time to track progress or lack of in successive physical therapy sessions.



Experimental results are shown in Figures 4.13 and 4.14, where $v_1$, $v_2$, and $v_3$ represent spine tilt, hip tilt, and shoulder tilt, respectively. Figure 4.13 compares two young females in 10-foot walking experiments. Three gait parameters are represented via box plots of 10 replicates. SE was used to summarize the variability of one walking path. The notation $W_{iv}$ indicates the distribution of 10 replicates for each experimental setting according to

$$W_{iv} \begin{cases} Subjects & i \in (1,2) \\ Parameters & v \in (1,2,3) \end{cases}$$

Sign of $\otimes$ indicates an average of SE values based on 10 replicates; the horizontal bar represents the median. The index v=1 means the spline tilt, v=2 is the hip tilt, and v=3 is the shoulder tilt.

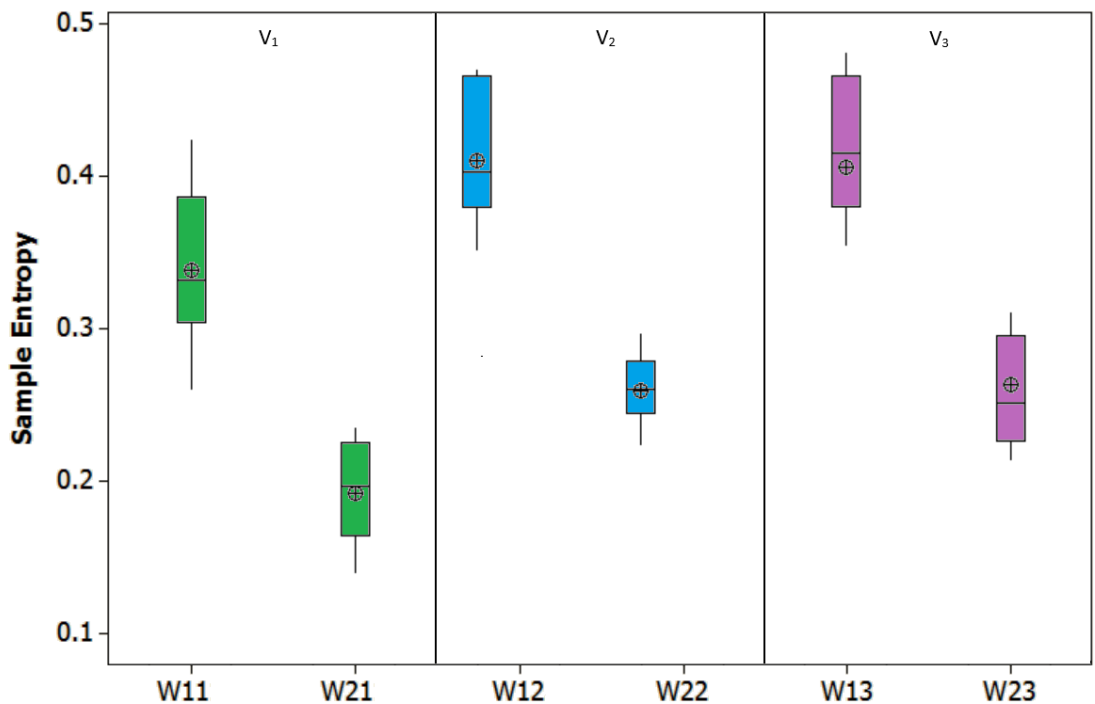

**Figure 4.13 Measure of variability using SE of three gait parameters on two subjects**

In order to compare the variabilities between the two subjects, $W_{1v}$ must be compared to $W_{2v}$. For example, SE values of subject 1 differed from SE values of subject 2 in terms of spine tilt



parameter ($V_1$), as shown in the first boxplot ($W_{11}$) and the second box plot ($W_{21}$) in Figure 4.13. Similarly, other pairs could also be compared. Results showed that subject 1 walked differently from subject 2.

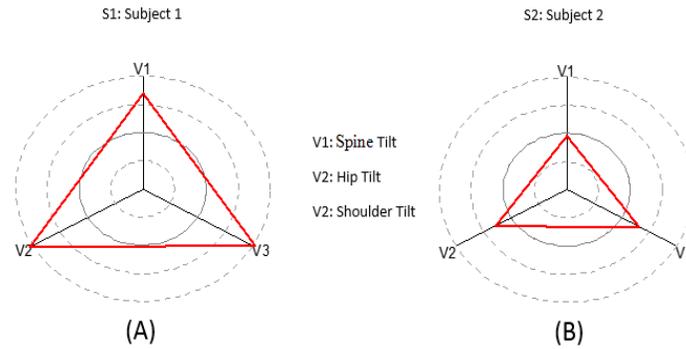

**Figure 4.14 Comparison of the mean SE values of the profiles using three gait parameters of Subject 1 (A) and Subject 2 (B) during walk experiments**

SE application to the time series demonstrated in (Richman and Moorman, 2000) could be expanded into multiple channels that track Kinect™ joints, as shown in Figure 3.5. Figure 4.14 compares movement profiles of the two subjects based on three gait parameters using star glyphs. This figure provides an efficient visual comparison to contrast differences between the two subjects (S1 and S2) using three gait parameters. It was evident that both subjects possess different gaits represented by the SE measures of three gait parameters.



## 4.3. Applying Machine Learning Classification Algorithms for Gait Analysis

This section explores the possibility of applying machine learning (ML) algorithms to discriminate an abnormal walking pattern from a normal pattern. Abnormal gait patterns are simulated by applying motion-restricting devices on subjects. Although the results presented in the previous section showed that the proposed method is capable of distinguishing human gaits, the analyses were based on one joint or one gait parameter at a time. ML algorithms would enable all joints to be considered at once and, therefore, provide better results. It would also help identify which joints have the most impact in distinguishing normal and abnormal walking conditions.

A new group of healthy students other than those who participated in a previous test (Explained in Section 3.1.1) to explore the accuracy and reliability of Kinect™, consisting of five females and five males as volunteers performed the 10-foot walk test in five different conditions for the new experiments. All participants declared no historical physical problems. They were asked to wear four different motion-restricting devices introduced in Section 3.1.2, which along with a normal walking make five distinct gaits and postural conditions. They performed three tentative trials for each condition and each subject which makes totally 150 test instances.

The first step in training various ML classification models is the preparation of input (independent) and output (dependent) variables. In this case study, the independent variables include the 15 joints mentioned in Section 3.2.2 while only dependent variable is the type of the motion-restricting devices. In this section, two classification approaches of the dependent variable are to be studied. The first approach separates all data into two classes – normal (30 observations) and abnormal walking (120 observations). The abnormal walking data were



simulated by asking subjects to perform the walking tests using the following motion-restricting devices: ankle brace, ACL brace, cane, and walker. The second approach retains five classes: normal walking, ankle brace, ACL brace, cane, and walker, each with 30 observations. The following section contains the analytical results for the first approach while those of the second approach will be discussed in section 4.3.2.

### 4.3.1 ML Study on Two-class Walking Classification

Four machine learning (ML) algorithms were selected for the classification problems, including logistic regression, random forests, K-nearest neighbors, and support vector machines. The binary logistic model is used to estimate the probability of a binary response based on one or more predictor or independent variables (features). It provides the estimate of odds ratio that allows the comparison of the two binary responses to be easily interpreted (Lemeshow & Hosmer, 1992). Random forests are a combination of tree predictors such that each tree depends on the values of a random vector sampled independently and with the same distribution for all trees in the forest (Breiman, 2001). The K-nearest neighbors rule is one of the oldest and simplest methods for pattern classification (Weinberger et al., 2006). Nevertheless, it often yields competitive results, and in certain domains, when cleverly combined with prior knowledge, it has significantly advanced the state-of the-art (Belongie et al., 2002). Support vector machine is a binary classification model. The binary classifier assumes that there are two classes in the task and each class is well identified by the decision surface (Wu et al., 2018).

The Weka data mining software (Version 3.8, 2018) and Python 3.6.4 (Python.org, 2018) were used to run these models based on 150 samples. A 10-fold cross-validation approach was



performed in that the test mode partitions the data into 90% training and 10% testing to build a model. Then this 10-fold approach repeats this validation tasks 10 times.

The default option in Weka was chosen for logistic regression and random forests algorithms. Oshiro et al. (2012) recommended a number of trees between 64 - 128 trees for the Random Forest algorithm. We chose to use the default number in Weka which is 100 trees for this algorithm. Also, the number of K-nearest neighbors was chosen based on the F-scores obtained for different K values. F-score, which is a standard measure, calculates the harmonic average of the precision and recall values as follows (Powers, 2007):

$$F = 2\frac{precision \cdot recall}{precision + recall} \tag{4.3}$$

where precision and recall are defined as (Olsen and Delen, 2008):

$$Precision = \frac{True\ positive}{True\ positive + False\ positive} \tag{4.4}$$

$$Recall = \frac{True\ positive}{True\ positive + False\ negative} \tag{4.5}$$

Thus, F-score results for different number of neighbors (K=1 to 7) were obtained in this case and K=1 was considered due to having the largest value of F-score (=0.93).

We did similar strategy for C value in Support Vector Machines algorithms. C is a trade-off between training error and the flatness of the solution. The larger the C value, the smaller the final training error will be (Bouboulis, 2012). However, increasing the C value too much may lead to lose the generalization properties of the classifier, because it will try to fit as best as possible



all the training points (including the possible errors of your dataset). In addition, a large C value usually increases the time needed for training (Guril & Anguri, 2009). A range of C values (1 to 10) were considered in the current study and the highest accuracy with the lowest errors obtained by C=5. As a result, K and C were considered as 1 and 5 to calculate accuracies of K-nearest Neighbors and Support Vector Machines algorithms, respectively. Figure 4.15 shows the summary of the results on the accuracy of the four applied ML algorithms in classifying 150 instances.

Our initial experimental results showed that truncating the data affects the accuracy of the classification algorithms. In fact, the truncated Kinect™ data improved the accuracy of all machine learning algorithms. Thus, prior to running the algorithms, all the extra points (joints dimensions) recorded while the studied subjects were standing, turning around, and anything but walking normally, were removed and the rest were extracted from the recorded files.

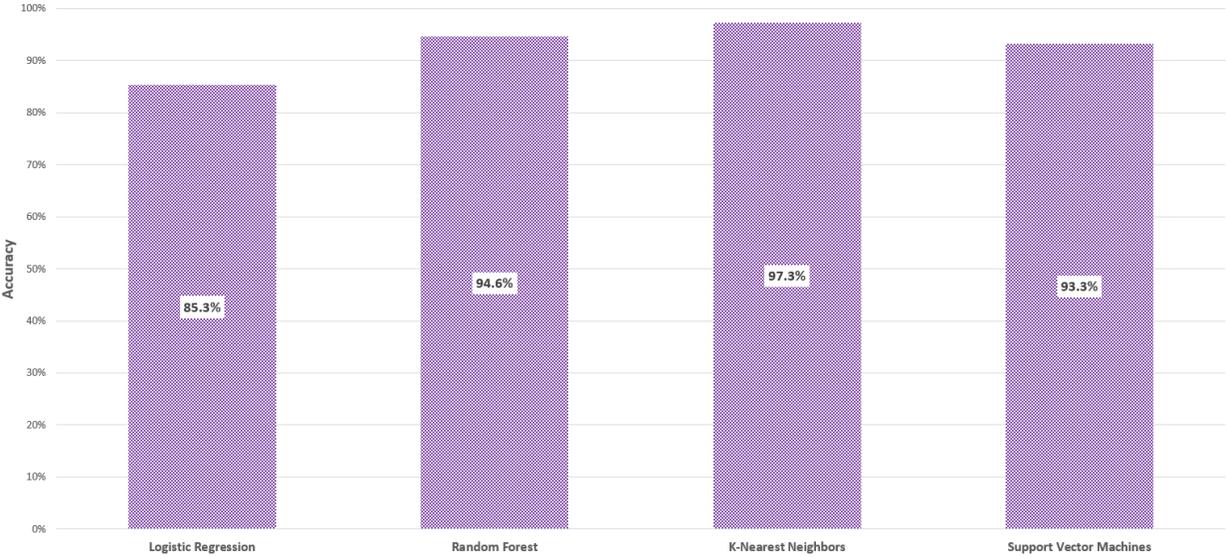

**Figure 4.15 The accuracy of four selected classification algorithms using the truncated data**



Figure 4.15 shows the accuracy results obtained for the four applied machine learning algorithms in classifying the test dataset correctly. As seen, the accuracy results from all selected algorithms are decent. Table 4.1 shows the confusion matrices obtained for all the studied algorithms.

**Table 4.1 Confusion Matrices for the accuracy results shown in Figure 4.15**

| Output Class | Logistic Regression | | Random Forests | | K-nearest Neighbors | | Support Vector Machines | | Classified as |
|---|---|---|---|---|---|---|---|---|---|
| | a | b | a | b | a | b | a | b | |
| | 20 | 10 | 25 | 5 | 26 | 4 | 23 | 7 | a = Normal Walk |
| | 12 | 108 | 3 | 117 | 0 | 120 | 3 | 117 | b = Abnormal Walk |
| | Target Class | | | | | | | | |

A hypothesis that we evaluated in this study, was to see whether dropping some attributes (Other than Device and Subject ID) affects the accuracy of the ML algorithms. Therefore, an attribute selection was run with the same data obtained from the normal walking trials in Python to find the most, moderate and least relevant joints to making predictions. A common practice is to keep the attributes with a moderate-to-high correlation and drop those attributes with a low correlation with the output variable. Dimension reduction reduces computational time.

The Correlation Attribute Evaluator (Hall, 1999) as a supervised technique was used with a ranker search method called Attribute Ranking, that evaluates the worth of each attribute on the full training data set by measuring the correlation (Pearson's) between it and the class and lists them in a rank order based on their significance. Table 4.2 shows this list obtained by individual evaluations of the attributes.



**Table 4.2 Attribute Selection Output**

| Attribute ID | Attribute Name | Ranked | Category |
|---|---|---|---|
| 13 | Left Shoulder | 0.50066 | A |
| 17 | Shoulder Spine | 0.48091 | A |
| 11 | Right Knee | 0.3777 | A |
| 8 | Left Hip | 0.3537 | A |
| 16 | Mid Spine | 0.30786 | A |
| 10 | Left Knee | 0.3072 | A |
| 9 | Right Hip | 0.30077 | A |
| 4 | Right Ankle | 0.24341 | B |
| 15 | Base Spine | 0.20313 | B |
| 14 | Right Shoulder | 0.1801 | B |
| 7 | Head | 0.17584 | B |
| 12 | Neck | 0.16796 | B |
| 5 | Left Foot | 0.12823 | C |
| 3 | Left Ankle | 0.10265 | C |
| 6 | Right Foot | 0.09725 | C |

By assigning the correlation 0.3 and 0.15 as the cut-off for relevant attributes, we could split all joints studied into three different categories in terms of impact. Group A represents the most significant joints, while the moderate and least significant joints are denoted by B and C, respectively. Then all the attributes in category C were removed from the analysis and the same classification algorithms were run on the remaining dataset to evaluate the defined hypothesis. Figure 4.16 indicates no changes for Random Forests and K-nearest Neighbors algorithms while logistic regression and support vector machines algorithms experienced a 6.7% increment and a 0.6% drop in prediction accuracy, respectively, after dropping category C from the accuracy studies. Table 4.3 also shows the confusion matrices obtained for the selected algorithms.



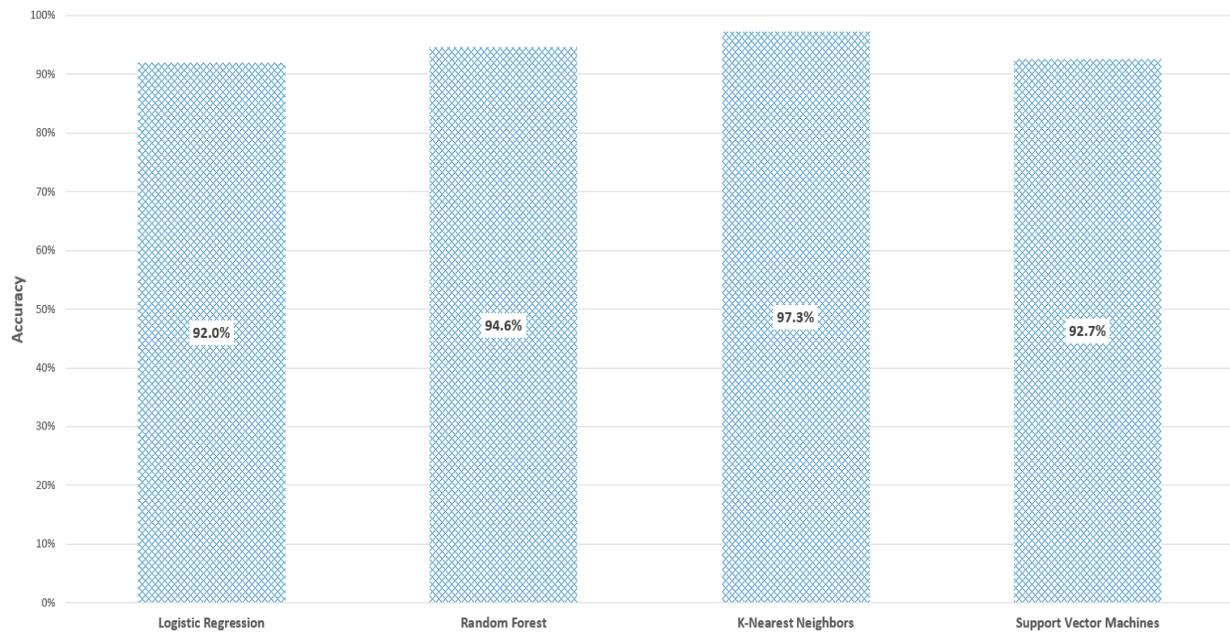

**Figure 4.16 The accuracy of four selected classification algorithms using the partial truncated data**

**Table 4.3 Confusion Matrices for the accuracy results shown in Figure 4.16**

| Output Class | Logistic Regression | | Random Forests | | K-nearest Neighbors | | Support Vector Machines | | Classified as |
|---|---|---|---|---|---|---|---|---|---|
| | a | b | a | b | a | b | a | b | |
| | 24 | 6 | 24 | 6 | 26 | 4 | 24 | 6 | a = Normal Walk |
| | 6 | 114 | 2 | 118 | 0 | 120 | 5 | 115 | b = Abnormal Walk |
| | Target Class | | | | | | | | |

Based on the analytical results, random forest and k-nearest neighbors algorithms showed the best potential to correctly differentiate an abnormal walking pattern from a normal walking pattern. Since the accuracy of these two algorithms were not changed after we removed category C from the study and also the dataset used in this case is not that large to affect the computational time, we decided to keep all fifteen joints as the features for the next approach.



### 4.3.2 ML Study on Multi-class Walking Classification

The next task in this study is to explore the abilities of machine learning algorithms in classifying all the four motion-restricting devices along with the normal patterns as the dependent variable by labeling the data into five different gait conditions: normal walking, ankle brace, ACL brace, cane, and walker. Logistical regression and support vector machine methods were no longer suitable since they were mainly used for binary class applications. Random forests and k-nearest neighbors were chosen as the two best classification algorithms on the truncated data to classify the data from all the fifteen joints into the five pre-defined groups. The data set is exactly the same as those introduced in the previous section.

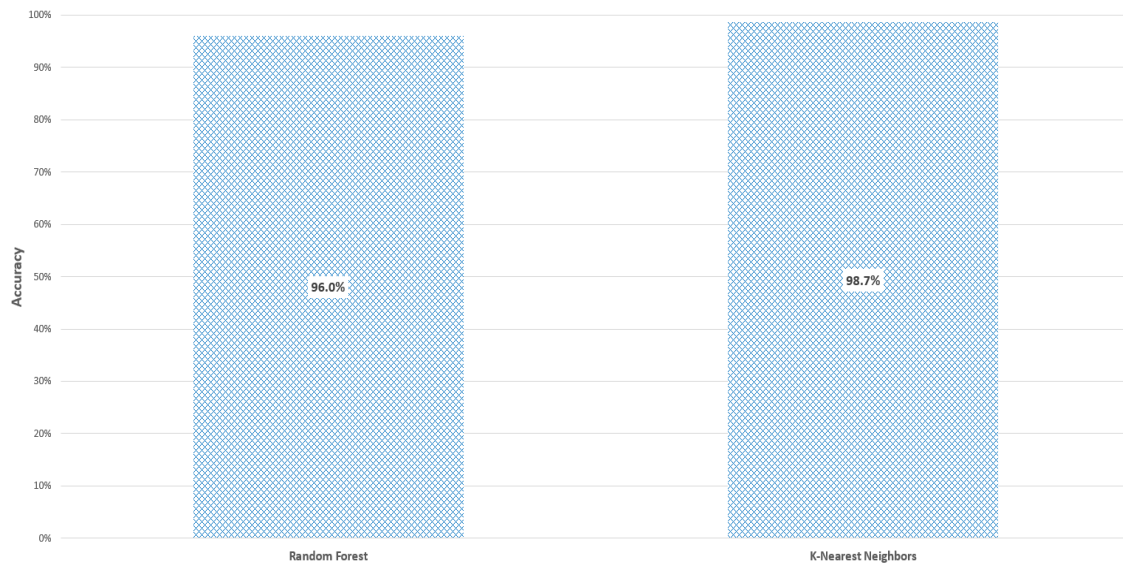

**Figure 4.17 The accuracy of two selected classification algorithms to differentiate the five gait conditions using the truncated data from the fifteen joints**

Figure 4.17 shows the performance of these methods in the data classification task into normal walking or the appropriate abnormal walking categories. Table 4.4 also shows the confusion



matrices obtained for the selected algorithms. These two algorithms showed higher accurate classification and fewer miss-classification rates when more classes of gait patterns were considered. We conjecture that the rand forests and K-nearest neighbors algorithms perform better when the training dataset is balance in that all classes contain 30 observations.

**Table 4.4 Confusion Matrices for the accuracy results shown in Figure 4.17**

| | Random Forests | | | | | K-nearest Neighbors | | | | | Classified as |
|---|---|---|---|---|---|---|---|---|---|---|---|
| | a | b | c | d | e | a | b | c | d | e | |
| Output Class | 30 | 0 | 0 | 0 | 0 | 30 | 0 | 0 | 0 | 0 | a = Normal Walk |
| | 0 | 27 | 2 | 0 | 1 | 0 | 30 | 0 | 0 | 0 | b = Cane |
| | 0 | 2 | 28 | 0 | 0 | 0 | 0 | 30 | 0 | 0 | c = Walker |
| | 0 | 0 | 1 | 29 | 0 | 0 | 0 | 1 | 29 | 0 | d = ACL |
| | 0 | 0 | 0 | 0 | 30 | 0 | 0 | 0 | 1 | 29 | e = Ankle |
| | Target Class | | | | | | | | | | |

The results of both implemented tasks are good, but they are based on a very small dataset. It is necessary to implement the same classification techniques on a larger number of samples to better evaluate properties of the proposed method.

Since the ML methods studied are able to detect possible changes in a person's walking pattern over time, with the proper training dataset, they may be applied to detect mental illness. For example, data set may be collected from people's walking data during a rehabilitation therapy after a stroke or recorded walking patterns of people with Parkinson, Multiple Sclerosis, and similar diseases. After the ML algorithms are trained, we might be able to predict possibilities of such a disease or even identify the severity level of these diseases. For example, our proposed technique can be a good alternative methodology for Sosnoff et al. (2012), Eltoukhy et al. (2017), and Tupa et al.'s (2015) studies.



# Chapter 5. Discussion and Conclusions

## 5.1 Summary of the Research

The series of conducted studies demonstrate the potential use of human kinematic measurements in clinical gait analysis. The proposed data analysis relies on a standardized experimentation process to gather gait data from Microsoft Kinect™ cameras and use the data to compare walking patterns of individuals. The proposed sample entropy measure was used to summarize the gait parameters or raw data from (X, Y, Z) format over time for each joint into one entropy value after each walking trial. Twenty-five SE values can be obtained so that gait changes can be identified.

Two main studies were conducted to (1) identify whether a person has deviated from his/her normal gaits by wearing motion-restricting devices used in everyday life, and (2) identify a distinct person based on his/her normal gaits. In Study 1, the experimental results analyzed by both SE values on gait parameters and on raw (X, Y, Z) data show that wearing the motion-restricting devices alters postural stability captured by 25 joints in healthy adults. Study 2 demonstrates that the proposed approaches are capable of distinguishing different persons using the proposed gait parameters. Moreover, the experimental results in Section 4.3 confirm that the proposed supervised learning methods are capable of classifying different walking patterns into normal walking and various abnormal walking patterns as one big class or individual classes. Multiple machine learning methods were applied to both studies. Although the accuracy results of the best two algorithms, shown in Figure 4.17 were close, K-nearest Neighbors method is the best for all tasks.



## 5.2 Applications and Discussions

The results in Study 1 show that a potential application of the proposed method may be for tracking the progress (or the lack of progress) in successive physical therapy sessions since the proposed method detects a person's normal and abnormal gaits. For example, it can help quantify, if something alters the postural status and diagnose if there is any disease such as Parkinson or Multiple Sclerosis that may alter gaits. The results of Study 2 show that the proposed methods can be used to distinguish different persons.

The experimental results demonstrate that wearing the motion-restricting devices alters walking patterns captured by relevant joints in healthy adults. The exact amount of changes can be quantified simply by using the proposed sample entropy measure (Shull et al., 2014). This study validates the hypothesis that the proposed personal profiles for individual subjects can be used to track changes in joints.

Moreover, the experimental results on different joints indicate that even a small number of joints are capable of identifying a person among 10 subjects. For a larger application of 100 or more persons, we conjecture that more parameters or joints may be necessary for proper identification.

Revised Sample Entropy (Chang et al., 2018) could also be applied to increase procedure sensitivity for detecting changes in both mean and variance. Moreover, additional test subjects are needed to improve the reliability of the conducted studies. Other statistical techniques may also be used in lieu of the SE since SE may not be familiar to most statisticians. The proposed



procedure in this thesis could be coupled with statistical process control techniques to monitor possible elevated fall risks of elderly people as well.

All systems have their limitations, ours is no different. The limitations of the proposed method stem mostly from depth camera's abilities. Like with all cameras we have restrictions based on lighting and camera angles. We tested in a well-lit room to make sure enough definition is available to the camera. As well as keeping the camera level at a set height to account for the errors that could have occurred. Our system works best when subjects wear shorts, tennis shoes, and a T-shirt. If a subject wears loose fitting pants the system returns data that is sporadic and therefore unusable. Also, the colors that subjects wear are important, black clothing usually ends up in unusable data. These limitations must be accounted for before we can expand the usage of the proposed system to other environments and situations.

## 5.3 Future Studies

Future studies may include elderly subjects as the target population. By tracking the walking profiles over time, it opens the doors to identify possible fall-related changes. It helps to measure any possible physical progression of an elderly with the help of our proposed sample entropy technique.

Some joints like hand-related joints were not included in the ML algorithm analysis, but may improve classification accuracy in another application. We envision that gaits from Multiple Sclerosis (MS) patients may be used to train an ML algorithm and then significant joints may be identified. This result may help detect or confirm the diagnosis of a patient who may have an early sign of MS symptoms through the proposed gait analysis.



Another potential future study may be a real-time identification system that can distinguish one subject from everyone else through his/her personal gait profile. We have studied the proposed system using 10 subjects. It is not clear what the dimensional restrictions (i.e. how many subjects) and the computational restriction (i.e. how fast the proposed method can be computed) are. This future study requires far more subjects to validate its results. Applying a big data computational method such as Map-reduce (Chu et al., 2007) may make the classification process much faster. Such a system can be used for monitoring people in hospitals and physical therapy centers. Finally, it is not clear how the proposed method can be implemented in a real-world as opposed to a controlled lab environment. More robust depth camera technologies may be necessary.

We are planning to implement this system in an environment that mirrors a small apartment. This testing would involve our current Kinect system as well as supplementing our data collection with CCTV.